\documentclass[preprint,authoryear,12pt]{elsarticle}
\usepackage{fullpage}
\usepackage[utf8]{inputenc} 
\usepackage{graphicx}
\usepackage{float}
\usepackage{hyperref}
\usepackage{subcaption}
\usepackage[export]{adjustbox}
\usepackage[bottom]{footmisc}
\usepackage{bm}
\usepackage[margin=2.5cm]{geometry}
\usepackage{amsmath,amsthm,amssymb}
\usepackage{textcomp}
\usepackage{graphicx,psfrag}
\newcommand{\be}{\begin{equation}}
\newcommand{\en}{\end{equation}}
\newcommand{\la}{\label}

\newcommand{\paa}{\partial}
\def\rr#1{(\ref{#1})}
\def\bm#1{\mbox{\boldmath{$#1$}}}
\def\ii{{\rm i}}

\numberwithin{equation}{section}
\renewcommand{\vec}[1]{\mathbf{#1}}
\usepackage[font=footnotesize]{caption}
\usepackage{titlesec}%set the format of the section title
\long\def\symbolfootnote[#1]#2{\begingroup\def\thefootnote{\fnsymbol{footnote}}\footnote[#1]{#2}\endgroup}
\def\rr#1{(\ref{#1})}
\def\bm#1{\mbox{\boldmath{$#1$}}}
\def\ii{{\rm i}}

\begin{document}
\begin{frontmatter}

\title{Localised bifurcation in soft cylindrical tubes under axial stretching and surface tension.}

\author[mymainaddress]{Dominic Emery}
\author[mymainaddress]{Yibin Fu\corref{mycorrespondingauthor}}
\cortext[mycorrespondingauthor]{Corresponding author}
\ead{y.fu@keele.ac.uk}
\address[mymainaddress]{School of Computing and Mathematics, Keele University, Staffordshire ST5 5BG, UK}

\begin{abstract}
\noindent We investigate localised bulging or necking in an incompressible, hyperelastic cylindrical tube under axial stretching and surface tension. Three cases are considered in which the tube is subjected to different constraints.
In case $1$ the inner and outer surfaces are traction-free and under surface tension, whilst in cases $2$ and $3$ the inner and outer surfaces (respectively) are fixed to prevent radial displacement and surface tension. However, each free surface in these latter two cases is still under surface tension. We first state the analytical bifurcation conditions for localisation and then validate them numerically whilst determining whether localisation is preferred over bifurcation into periodic modes.
It is shown that bifurcation into a localised solution is unattainable in case $1$ but possible and favourable in cases $2$ and $3$. In contrast, in case $1$ any bifurcation must necessarily take the form of a periodic mode with a non-zero wave number. Our results are validated using Finite Element Method (FEM) simulations.
\end{abstract}


\end{frontmatter}
\section{Introduction}
In fluid mechanics, surface tension is the architect of many beautiful phenomena such as water walking insects and the bundling of wetted lamellae \citep{bico2004,bush2006}. Perhaps the most famous is the Rayleigh-Plateau instability \citep{plateau1873,rayleigh1892}, which is manifested in the destabilisation of uniform cylindrical columns of viscous fluid into a succession of droplets. With the surface of the fluid acting essentially as a stretched membrane due to intermolecular forces, the desire to reduce the surface area to volume ratio causes a transformation into spherical droplets, thus lowering the total energy in tandem \citep{degennes2013}.

In recent years, interest in surface tension has transcended into the field of solid mechanics \citep{lf2012}. Whilst surface tension is negligible in the classic continuum framework above the elastocapillary length scale $\gamma/\mu$ \citep{style,bico2018} (where $\gamma$ is the surface tension and $\mu$ the shear modulus), it dominates bulk elastic forces in microscale soft materials such as gels and biological tissue. Thus, the development of the field of elastocapillarity has become a necessity in understanding surface instabilities in swollen hydrogels \citep{tanaka1992} and soft substrates under mechanical constraints \citep{mora2011, chen, ciarletta2014wrinkle}, for example. Moreover, surface tension has been shown recently to influence genetic diseases such as \textit{lissencephaly}, which is characterised by a reduction of sulci in brain organoids \citep{dobyns1993,engstrom2018}. By modelling said organoids as a soft solid cylinder encapsulated by a growing elastic layer, \cite{riccobelli2020} showed that reduction in tissue stiffness (and thus increased elasto-capillary effects) stabilised the tube against circumferential buckling modes, thus offering a theoretical explanation for lissencephaly.

Peristaltic instability in soft cylinders, commonly referred to as ``beading" or ``pearling", has been implicated in axonal degeneration due to cytoskeletal trauma \citep{kilinc,goriely2015} and neurodegenerative disorders such as Alzheimer's and Parkinson's disease \citep{datar2019}. Thus, a host of studies have attempted to resolve the theoretical perplexities surrounding this phenomenon. In the case of a solid cylinder, contributions come from \citet{bs1996}, \citet{bc2003}, \citet{mp2010}, \citet{cb2012}, \cite{taffetani}, and \cite{xuan2016}, all of whom conclude that beading is a long wavelength instability. Only very recently has the explicit nature of this localised solution become apparent. Both \cite{xuan2017} and \cite{giudici2020} showed that beading is in fact a phase separation phenomenon, whilst \cite{FuST} focused on the characterisation of localised solutions that can bifurcate from the uniform state and demonstrated that a variety of localised bifurcation behaviours such as necking and bulging can exist depending on the loading path.

%under fixed axial stretch $\lambda$ and increasing surface tension, a minimum critical surface tension occurs for some fixed $\lambda_{\text{min}}$ such that for $\lambda <\lambda_{\text{min}}$ ($\lambda >\lambda_{\text{min}}$) the localised solution corresponds to a depression (bulge).

Localised bulging has been extensively studied in hollow tubes under internal inflation and without surface tension; see, for example, \cite{chater1984}, \cite{kyriakides1991}, \cite{fu2008}, \cite{alhayani2014competition}, \cite{fu2016localized} and the references therein. However, when inflation is forgone and surface tension effects are introduced, theoretical works are far less concrete. \cite{henann} conducted FEM simulations for bifurcation from finitely deformed tubes which are externally or internally fixed, whilst \cite{xuan2016} proposed analytically that the bifurcation of a cylindrical cavity in an infinite incompressible solid is again associated with infinite wavelength. Most recently, \cite{liuwang} examined two of the three cases alluded to in the Abstract. Surprisingly, an analytical solution was obtained for the governing equation. This was contrary to expectations given the investigations of \cite{HO1979} into tubes under axial tension and internal pressure, whose boundary value problem could only be solved numerically. There is clearly a need to resolve this discrepancy and to deduce absolutely whether localised bifurcation can occur in cylindrical tubes under different constraints. It turns out that the predictions based on our current analysis are quite different from those given by \cite{liuwang}. For instance, for case 1 \cite{liuwang} predicted that the critical wavelength is also infinite, but our analysis shows that such a mode is associated with negative values of surface tension and therefore cannot physically occur.

The remainder of this paper is divided into five sections as follows. After formulating the problem in section $2$, we present in section $3$ analytical conditions for localised bifurcation by drawing upon known results for the analogous problem of localised bulging of inflated hyperelastic tubes. In section $4$ we firstly elaborate on the need for further analysis of the problem at hand. We then conduct a linear bifurcation analysis for each of the three cases under consideration, and produce a numerical relationship between the bifurcation parameter and the wave number. Based on this relationship we deduce for each case whether a localised solution can bifurcate from the finitely deformed state and, where it can, we determine conditions for localised bifurcation numerically via a determinant shooting method. In section 5 we conduct FEM simulations to validate our theoretical predictions. Finally, concluding remarks are offered in section $6$.

%%%%%%%%%%%%%%%%%%%%%%%%%%%%%%%%%%%%%%%%%%%%%%%%%%%%%%%%%%%%%%%%%%%%%%%%%%%%%%%%%%%%%%
%%%%%%%%%%%%%%%%%%%%%%%%%%%%%%%%%%%%%%%%%%%%%%%%%%%%%%%%%%%%%%%%%%%%%%%%%%%%%%%%%%%%%%

\section{Problem formulation}
Consider a hyperelastic cylindrical tube whose reference configuration $\mathcal{B}_0$ is defined in terms of the cylindrical polar coordinates $\left(R,\,\Theta,\,Z\right)$ such that
\begin{align}
A\leq R\leq B,\,\,\,\,\,\,\,\,0\leq\Theta\leq 2\pi&,\,\,\,\,\,\,\,\,-L\leq Z\leq L, \label{tubedimensions}
\end{align}
where $A$ and $B$ are respectively the undeformed inner and outer radii and the cylinder has an axial half-length $L$. The position vectors of a representative material particle in the reference configuration $\mathcal{B}_0$ and the finitely deformed configuration $\mathcal{B}_e$ are denoted $\vec{X}$ and $\vec{x}$ respectively, such that
\begin{align}
\vec{X}&=R\,\vec{E}_R + Z\,\vec{E}_Z,\,\,\,\,\,\,\,\,\vec{x}=r\,\vec{e}_r + z\,\vec{e}_z, \label{particlecoords}
\end{align}
where $\left(r,\,\theta,\,z\right)$ are the coordinates of $\vec{x}$ and $\left(\vec{E}_R,\vec{E}_{\Theta},\vec{E}_Z\right)$ and $\left(\vec{e}_r,\vec{e}_{\theta},\vec{e}_z\right)$ are the orthonormal bases of $\mathcal{B}_0$ and $\mathcal{B}_e$, respectively. For the sake of generality, we assume for the meantime that both the inner and outer surfaces are unconstrained, and thus we denote by $a$ and $b$ the inner and outer radii in $\mathcal{B}_e$. A general axi-symmetric deformation of the tube can be characterised by the following variable transformations
\begin{align}
r&=r\left(R,Z\right),\,\,\,\,\,\,\,\,\theta=\Theta,\,\,\,\,\,\,\,\,z=z\left(R,Z\right). \label{gendef}
\end{align}
The deformation gradient $\vec{F}$ is defined by $d \vec{x}=\vec{F} d\vec{X}$ and  takes the following form:
\begin{align}
\vec{F}&=\frac{\partial r}{\partial R}\,\vec{e}_r\otimes\vec{E}_R+\frac{\partial r}{\partial Z}\,\vec{e}_r\otimes\vec{E}_Z+\frac{r}{R}\,\vec{e}_\theta\otimes\vec{E}_{\theta}+\frac{\partial z}{\partial R}\,\vec{e}_z\otimes\vec{E}_R+\frac{\partial z}{\partial Z}\,\vec{e}_z\otimes\vec{E}_Z. \label{Fgen}
\end{align}
The cylindrical tube is assumed to be incompressible, and so the following constraint of isochorism must be satisfied
\begin{align}
\text{det}\,\vec{F}&=1. \label{detF}
\end{align}
For the sake of simplicity, we assume that the constitutive behaviour of the tube is governed by a strain energy function of the form
\begin{align}
W&=W\left(I_B\right), \label{SEfunction}
\end{align}
where $I_B$ is the first principal invariant of the left Cauchy-Green strain tensor $\vec{B}=\vec{F}\vec{F}^\top$, i.e. $I_B=\text{tr}\,\vec{B}$ and the superscript $\top$ denotes transposition. This form of the strain energy function includes neo-Hookean and Gent material models as special cases, and there is some evidence that it is capable of giving results that are at least qualitatively correct for the kind of deformation under consideration \citep{wineman2005, zhou2018}. To simplify presentation, we shall only present our analytical results for the case where the tube material is neo-Hookean. However, our actual derivations are carried out with the aid of Mathematica \citep{wo2019} for the general strain energy \rr{SEfunction} allowing for some results to be presented for the Gent material model when comparison is made with FEM simulations. The neo-Hookean and Gent material models are given respectively as follows:
\begin{equation}
W(I_B)=\frac{1}{2}\,\mu\left(I_B -3\right), \;\;\;\;\;\;\;\; W(I_B)=-\frac{1}{2}\,J_m\,\mu\,\log\left(1-\frac{I_B -3}{J_m}\right), \label{neohook}
\end{equation}
where $\mu$ is the shear modulus and $J_m$ is the extensibility limit.  In the limit $J_m\rightarrow\infty$, the neo-Hookean strain energy function $(\ref{neohook})_1$ is recovered from the Gent model. For the remainder of this paper we scale all lengths by $B$ and stresses by $\mu$. Therefore, we may set $B=1$ and $\mu=1$ without loss of generality.

\subsection{Stream-function formulation}

As proposed by \citet{ciarletta2011}, we may consider a reformulation of the problem in terms of the mixed co-ordinate stream function $\phi=\phi\left(R,z\right)$ which enforces the incompressibility constraint $(\ref{detF})$ exactly through the relations
\begin{align}
 r^2&=2\,\phi_{,z},\,\,\,\,\,\,\,\,Z=\frac{1}{R}\,\phi_{,R}, \label{incphi}
\end{align}
where a comma denotes partial differentiation with respect to the implied coordinate. Now, $(\ref{incphi})$ may be applied in conjunction with the chain rule to re-express $\vec{F}$ in terms of $\phi$ as such
\begin{align}
\vec{F}&=\frac{\left[\phi_{,Rz}+\frac{\phi_{,zz}}{\phi_{,Rz}}\left(\frac{\phi_{,R}}{R}-\phi_{,RR}\right)\right]}{\sqrt{2\,\phi_{,z}}}\,\vec{e}_r\otimes\vec{E}_R + \frac{R\,\phi_{,zz}}{\sqrt{2\,\phi_{,z}}\,\phi_{,Rz}}\,\vec{e}_r\otimes\vec{E}_Z + \frac{\sqrt{2\,\phi_{,z}}}{R}\,\vec{e}_{\theta}\otimes\vec{E}_{\Theta} \nonumber\\[1em]
&\,\,\,\,\,\,\,+\frac{\left[\frac{\phi_{,R}}{R}-\phi_{,RR}\right]}{\phi_{,Rz}}\,\vec{e}_z\otimes\vec{E}_R + \frac{R}{\phi_{,Rz}}\,\vec{e}_z\otimes\vec{E}_Z. \label{Fphi}
\end{align}
Thus, $I_B$ is determined to take the form
\begin{align}
I_B=\frac{\left[\phi_{,Rz}-\frac{R\,\phi_{,zz}}{\phi_{,Rz}}\left(\frac{\phi_{,RR}}{R}-\frac{\phi_{,R}}{R^2}\right)\right]^2}{2\,\phi_{,z}}+\frac{1}{2}\frac{R^2\,\phi_{,zz}^2}{\phi_{,z}\,\phi_{Rz}^2}+\frac{2\,\phi_{,z}}{R^2}+\frac{R^2}{\phi_{,Rz}}+\frac{R^2}{\phi_{,Rz}^2}\left[\frac{\phi_{,R}}{R^2}-\frac{\phi_{,RR}}{R}\right]^2. \label{IBphi}
\end{align}

A variational approach is considered in deriving the equilibrium equation and the associated boundary conditions. We introduce the total potential energy $\mathcal{E}$ which comprises of the bulk elastic energy $\mathcal{E}_b$ and the surface energies $\mathcal{E}^{A}_{s}$ and $\mathcal{E}_{s}^{B}$ on the inner and outer boundaries such that
\begin{align}
\mathcal{E}&=\mathcal{E}_b + \mathcal{E}^{A}_{s} + \mathcal{E}^{B}_{s}, \label{E}
\end{align}
where $\mathcal{E}_b$, $\mathcal{E}^{A}_{s}$ and $\mathcal{E}_s^B$ are given in terms of $\phi$ and its partial derivatives as follows
\begin{align}
\mathcal{E}_b&=2\,\pi\int^{\ell}_{-\ell}\int^{B}_{A}\,\phi_{,Rz}\,W(I_B)\,dR\,dz,\,\,\,\,\,\,\,\,\mathcal{E}_{s}^{A,\,B}=2\,\pi\,\gamma\int^{\ell}_{-\ell}\,
\left.\sqrt{2\phi_{,z}+\phi_{,zz}^{2}}\right\vert_{R=A,\,B}dz. \label{Ebphi}
\end{align}
In the above expression, $\gamma$ denote the surface tension scaled by $\mu B$ and $\ell$ the axial half-length in $\mathcal{B}_e$. Note that in $(\ref{Ebphi})_1$ $I_B$ is given by $(\ref{IBphi})$ and use has been made of the relation $dZ=\frac{\partial Z}{\partial z}dz=\frac{1}{R}\,\phi_{,Rz}\,dz$.
The equilibrium equation corresponds to the vanishing of the first variation of $(\ref{Ebphi})_1$. Equivalently, we must solve the Euler-Lagrange equation
\begin{align}
\left(\frac{\partial \mathcal{L}_b}{\partial \phi_{,iA}}\right)_{,iA}-\left(\frac{\partial \mathcal{L}_b}{\partial \phi_{,j}}\right)_{,j}&=0, \label{goveqn}
\end{align}
where the standard summation convention is applied, with $j=R$ or $z$ and $iA = RR$, $Rz$ or $zz$, and the bulk Lagrangian $\mathcal{L}_b$ is defined by
\begin{align}
\mathcal{L}_{b}&=\phi_{,Rz}\,W(I_B). \label{bulklagrangian}
\end{align}
In case 1, the curved surfaces $R=A$ and $R=B$ are traction-free and under surface tension, and these boundary conditions take the respective forms
\begin{align}
\left[\frac{\partial \mathcal{L}_b}{\partial \phi_{,R}}-\left(\frac{\partial \mathcal{L}_b}{\partial \phi_{,RR}}\right)_{,R}-\left(\frac{\partial \mathcal{L}_b}{\partial \phi_{,Rz}}\right)_{,z}\right]_{R=A}-\left(\frac{\partial \mathcal{L}_s^{A}}{\partial \phi_{,zz}}\right)_{,zz}+\left(\frac{\partial \mathcal{L}_s^{A}}{\partial \phi_{,z}}\right)_{,z}&=0, \label{BC1A}
\end{align}
\begin{align}
\left[\frac{\partial \mathcal{L}_b}{\partial \phi_{,R}}-\left(\frac{\partial \mathcal{L}_b}{\partial \phi_{,RR}}\right)_{,R}-\left(\frac{\partial \mathcal{L}_b}{\partial \phi_{,Rz}}\right)_{,z}\right]_{R=B}+\left(\frac{\partial \mathcal{L}_s^{B}}{\partial \phi_{,zz}}\right)_{,zz}-\left(\frac{\partial \mathcal{L}_s^{B}}{\partial \phi_{,z}}\right)_{,z}&=0, \label{BC1B}
\end{align}
where the inner and outer surface Lagrangian's $\mathcal{L}_s^A$ and $\mathcal{L}_s^B$ are defined by
\begin{align}
\mathcal{L}_s^{A,\,B}&=\gamma\left.\sqrt{2\,\phi_{,z}+\phi_{,zz}^{2}}\right\vert_{R=A,\,B}. \label{surflagrangianA}
\end{align}
It is noted that the opposite signs of the surface Lagrangian terms in $(\ref{BC1A})$ and $(\ref{BC1B})$ signify the opposing mean curvatures of the inner and outer surfaces. In cases 2 and 3, the inner and outer surfaces (respectively) are constrained to prevent radial displacement, with the other curved boundary remaining traction-free. In these circumstances, we require that the incremental radial displacement on the fixed surface vanishes. For all three cases, we have zero shear forces on $R=A$ and $R=B$, invoking two further boundary conditions which are expressed as follows
\begin{align}
\left.\frac{\partial \mathcal{L}_b}{\partial \phi_{,RR}}\right\vert_{R=A,B}&=0.  \label{BC2}
\end{align}
%To simplify presentation, we shall only present our analytical results for the case where the tube material is neo-Hookean. However, our actual derivations are carried out with the aid of Mathematica \citep{wo2019} for the general strain energy \rr{SEfunction} allowing for some results to be presented for the Gent material model when comparison is made with FEM simulations. The neo-Hookean and Gent material models are given respectively as follows
%\begin{equation}
%W(I_B)=\frac{1}{2}\,\mu\left(I_B -3\right), \;\;\;\;\;\;\;\; W(I_B)=-\frac{1}{2}\,J_m\,\mu\,\log\left(1-\frac{I_B -3}{J_m}\right), \label{neohook}
%\end{equation}
%where $\mu$ is the shear modulus and $J_m$ is the extensibility limit.  In the limit $J_m\rightarrow\infty$, the neo-Hookean strain energy function $(\ref{neohook})_1$ is recovered from the Gent model. For the remainder of this paper we scale all lengths by $B$, stresses by $\mu$ and $\gamma$ by $\mu B$. Therefore, we may set without loss of generality $B=1$ and $\mu=1$.
\section{The primary deformation and bifurcation conditions for localisation}
\setcounter{equation}{0}
We first characterize the following primary axi-symmetric deformation, a sub-class of $(\ref{gendef})$,  that is theoretically possible for all values of surface tension $\gamma$ and principal axial stretches $\lambda=\ell/L$
\begin{align}
r=r(R),\,\,\,\,\,\,\,\,\theta=\Theta,\,\,\,\,\,\,\,\,z=\lambda Z. \label{axdef}
\end{align}
The associated deformation gradient is given by
\begin{align}
\vec{F}&=\frac{\partial r}{\partial R}\,\vec{e}_r\otimes\vec{E}_R + \frac{r}{R}\,\vec{e}_{\theta}\otimes\vec{E}_{\Theta}+\lambda\,\vec{e}_z\otimes\vec{E}_Z. \label{Fax}
\end{align}
Upon substituting $(\ref{Fax})$ into $(\ref{detF})$ and integrating the resulting equation, we obtain
\begin{align}
r(R)&=\sqrt{\lambda^{-1}\left(R^2-A^2\right)+a^2}. \label{rR}
\end{align}
Through further integration of $(\ref{incphi})$, we find that the corresponding stream function, denoted by $\phi_0$, takes the form
\begin{align}
\phi_0&=\frac{R^2 z}{2\,\lambda}+\frac{1}{2}\left(a^2-\frac{A^2}{\lambda}\right)z. \label{phi0gen}
\end{align}
The outer deformed radius is defined from $(\ref{rR})$ as $b=\sqrt{\lambda^{-1}\left(B^2-A^2\right)+a^2}$. We consider the three cases alluded to in the Abstract separately.

%
%We firstly demonstrate that bifurcation of soft tubes from the trivial state governed by $(\ref{phi0gen})$ into a localised non-trivial solution has a clear physical interpretation. This is achieved by seeking a localised bifurcation condition analytically from the variational framework introduced in section $2$.

%%%%%%%%%%%%%%%%%%%%%%%%%%%%%%%%%%%%%%%%%%%%%%%%%%%%%%%%%%%%%%%%%%%%%%%%%%%%%%%%%%%%%%
%%%%%%%%%%%%%%%%%%%%%%%%%%%%%%%%%%%%%%%%%%%%%%%%%%%%%%%%%%%%%%%%%%%%%%%%%%%%%%%%%%%%%%

\subsection*{Case 1: Traction-free curved boundaries under surface tension}
We first consider the case whereby the inner and outer surfaces of the tube are traction-free and under surface tension. Under these conditions, the inner deformed radius $a$ is an unknown quantity. We assume that the tube is subject to the combined action of surface tension and a resultant axial force $\mathcal{N}$, which modifies the total potential energy $(\ref{E})$ as follows
\begin{align}
\mathcal{E}=\mathcal{E}_b+\mathcal{E}_s^A+\mathcal{E}_s^B-\left(\lambda -1\right) \mathcal{N}. \label{TPEcase2}
\end{align}
For the primary deformation, $\mathcal{E}$ can be evaluated by substituting $(\ref{phi0gen})$ into $(\ref{Ebphi})$. To satisfy equilibrium, we  require that
$\paa \mathcal{E}/\paa a=0$ and  $\paa \mathcal{E}/\paa \lambda =0$. Corresponding to the neo-Hookean material model, the former yields an equation for $\gamma=\gamma\left(\lambda,\,a\right)$ as follows
\begin{equation}
\gamma=\frac{(a^2\,\lambda - A^2) (a-b)}{2\,a\,b\,\lambda^2}+\frac{a\,b}{ \lambda\,(a+b) } \log \left(\frac{Ab}{a}\right),
 \label{gamcase1}
\end{equation}
whereas the latter gives the following expression for
 $\mathcal{N}=\mathcal{N}(\lambda,\,a)$
%\begin{align}
%\mathcal{N}&=\frac{\pi}{2} \Bigg[\Bigg.4\,\gamma\,\lambda^2\,b^2\left(a+b\right)+\lambda\,b\,\left(2\,\gamma-2\,\lambda^2\,b\right)\left(A^2 -1\right)+b^2\left(a^2\,\lambda +A^2 -2\right)+a^2\,\left(A^2 -a^2\,\lambda\right)\nonumber\\[1em]
%&\,\,\,\,\,\,\,+2\,A^2\,b^2\,\log\left(\frac{a}{A\,b}\right)\Bigg.\Bigg]. \label{axforcecase1}
%\end{align}
\begin{align}
\mathcal{N}=\frac{\pi}{2\,\lambda^2} \Bigg[\Bigg.4\,a\,\gamma+\frac{a^2}{b^2}\left(A^2 -2\right)\left(\lambda^3 -1\right)+\frac{2\,\gamma\,\lambda}{b}\left(a^2+b^2\right)+2\,A^2\log\left(\frac{a}{A\,b}\right)\Bigg.\Bigg], \label{axforcecase1}
\end{align}
with $\gamma$ eliminated through substitution of $(\ref{gamcase1})$. Alternatively, \rr{gamcase1} can be derived with the aid of the Cauchy stress tensor $\sigma$, defined through the constitutive equation $\sigma=2\,W'\left(I_B\right)\,\vec{B}-p\,\vec{I}$, together with the boundary conditions $\sigma_{rr}|_{r=a}=\gamma/a$ and $\sigma_{rr}|_{r=b}=-\gamma/b$. where $p$ is the Lagrangian multiplier associated with the constraint of incompressibility and $\vec{I}$ is the identity tensor. $\mathcal{N}$ is then equal to the resultant of  $\sigma_{zz}$ plus $2\,\pi\,\gamma\,(a+b)$.

Mathematically, the above two relations
 $\gamma=\gamma\left(\lambda,\,a\right)$ and $\mathcal{N}=\mathcal{N}\left(\lambda,\,a\right)$ cannot be inverted to express $\lambda$ and $a$ uniquely in terms of $\gamma$ and $\mathcal{N}$ when
 \begin{align}
\mathcal{J}\left(\gamma,\,\mathcal{N}\right)&\equiv \frac{\partial \gamma}{\partial\,a}\frac{\partial\mathcal{N}}{\partial \lambda}-\frac{\partial \gamma}{\partial \lambda}\frac{\partial \mathcal{N}}{\partial\,a}=0, \label{jacobian}
\end{align}
where $\mathcal{J}(\gamma,\,\mathcal{N})$ is the Jacobian of the vector function $(\gamma,\,\mathcal{N})$. Based on the analysis of \cite{fu2016localized}, we may conjecture that this is the condition for localisation. It will be verified in the next section that this is the condition for a bifurcation mode with zero axial wave number to exist. Alternatively, this is the condition for zero to become a triple eigenvalue of a certain spectral eigenvalue problem governing the incremental perturbations \citep{kirch, Iooss}. When $a$ or $b$ is fixed (cases 2 and 3 to be discussed shortly), the above bifurcation condition reduces to $\paa \mathcal{N}/\paa \lambda =0$ for fixed surface tension or $\partial \gamma/\partial \lambda =0$ for fixed axial force.

\begin{figure}[h!]
\begin{center}
\includegraphics[scale=0.55]{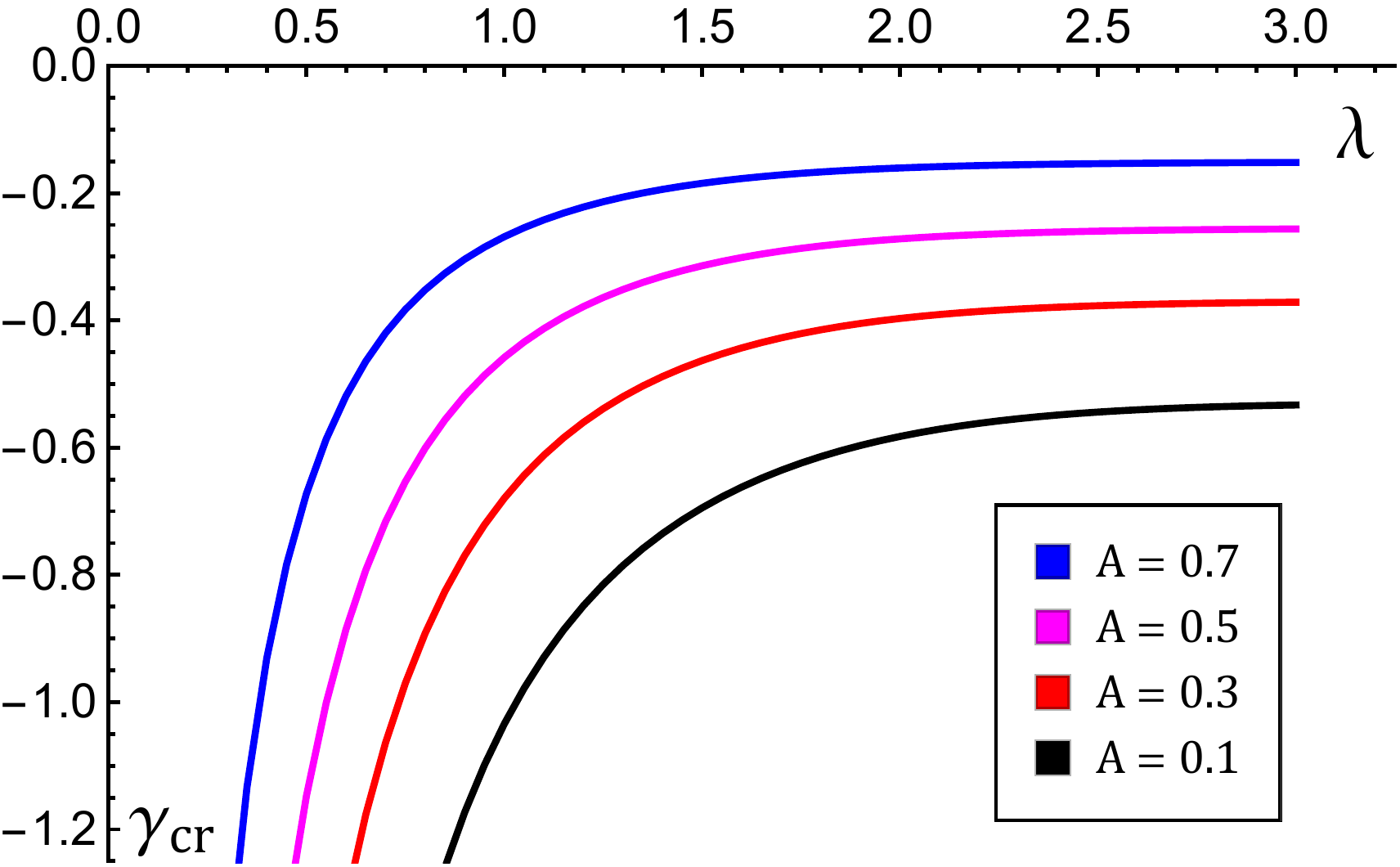}
\vspace{3mm}
\caption{Solutions of the bifurcation condition \rr{jacobian} for localisation in the $\left(\lambda,\,\gamma_{cr}\right)$ plane. The vertical order of the curves and the parameter values in the legend are equivalent.}
\label{fig1}
\end{center}
\end{figure}

 Fig. $\ref{fig1}$ shows contour plots in the $(\lambda, \gamma_{cr})$ plane of the bifurcation condition $(\ref{jacobian})$ for four typical values of $A$. It is seen that the critical surface tension $\gamma_{cr}$ is always negative, which suggests that localisation is not possible in this case. This will be confirmed in the next section where we also show that bifurcation into periodic modes are possible provided $\mathcal{N}$ is negative and has sufficient magnitude.
 %However, at present this statement is only a postulation and must be concretely verified by deducing the relationship between the chosen loading parameter and the mode number $k$ of the non-trivial solution. Indeed, for bifurcation into a localised solution, we expect that the critical value of the loading parameter will correspond to the mode number $k_{cr}=0$ (\cite{kirch}, \cite{Iooss}). These queries are investigated through linear bifurcation analysis further into this work.

%%%%%%%%%%%%%%%%%%%%%%%%%%%%%%%%%%%%%%%%%%%%%%%%%%%%%%%%%%%%%%%%%%%%%%%%%%%%%%%%%%%%%%
%%%%%%%%%%%%%%%%%%%%%%%%%%%%%%%%%%%%%%%%%%%%%%%%%%%%%%%%%%%%%%%%%%%%%%%%%%%%%%%%%%%%%%

\subsection*{Case 2: Radially fixed inner boundary free of surface tension}
\begin{figure}[h!]
\centering
\begin{subfigure}[t]{0.486\textwidth}
\includegraphics[width=\linewidth, valign=t]{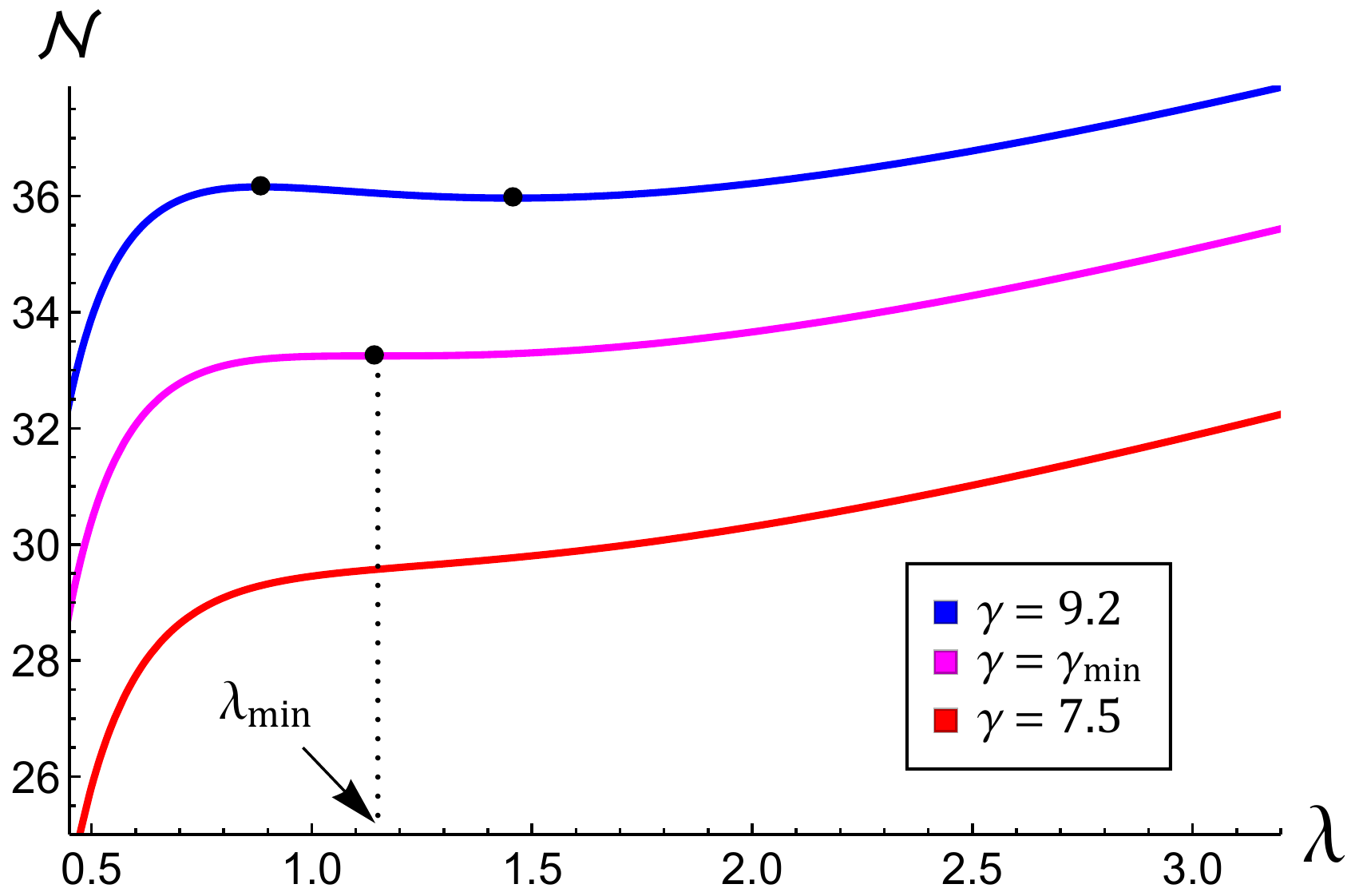}
\subcaption*{\textbf{(a)}}
\end{subfigure}\hfill
\begin{subfigure}[t]{0.483\textwidth}
\includegraphics[width=\linewidth, valign=t]{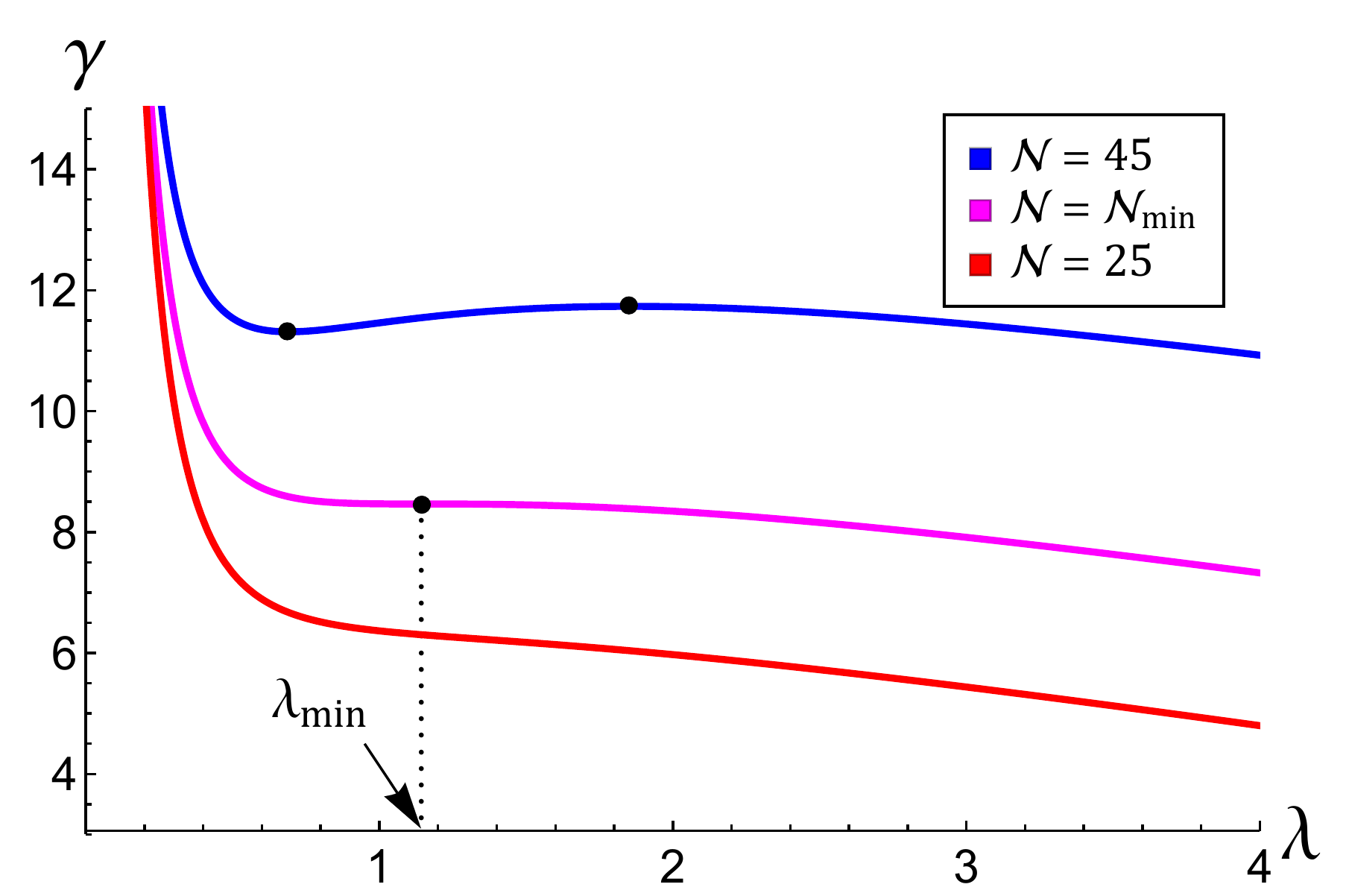}
\subcaption*{\textbf{(b)}}
\end{subfigure}
\caption{\textbf{(a)} The variation of $\mathcal{N}$ with respect to $\lambda$ for a tube of inner radius $A=0.5$ under several fixed surface tensions. As $\gamma$ is increased from zero, $\mathcal{N}$ is initially a monotonically increasing function of $\lambda$ until $\gamma$ reaches $\gamma_{\text{min}}$, after which $\mathcal{N}$ has a maximum and a minimum.  \textbf{(b)} The variation of $\gamma$ with respect to $\lambda$ for various fixed $\mathcal{N}$ where $A=0.5$. Only when $\mathcal{N}$ becomes larger than $\mathcal{N}_{\text{min}}$ will the variation become non-monotonic and localized bifurcation become possible. The vertical order of curves and legend parameter values are equivalent.}
\label{fig2}
\end{figure}

In case 2, prevention of radial displacement of the inner surface requires we fix $a=A$, and the absence of surface tension on this boundary means that $\mathcal{E}_s^A=0$. Thus, $\phi_0$ and $b$ become
\begin{align}
\phi_0&=\frac{R^2\,z}{2\,\lambda}+\frac{A^2}{2}\left(1 -\frac{1}{\lambda}\right)\,z, \;\;\;\; b=\sqrt{\lambda^{-1} (1-A^2)+A^2}. \label{phi0case2}
\end{align}
In this case, the single parameter $\lambda$ is sufficient to determine the deformation completely. Therefore, equilibrium requires only that $\partial \mathcal{E}/\partial \lambda=0$, from which we obtain the following expression for $\mathcal{N}=\mathcal{N}(\lambda)$ where $\gamma$ is fixed
\begin{align}
\mathcal{N}&=\frac{1}{\pi}\Bigg[\Bigg.\frac{\left(1-\lambda\right)}{2\,\lambda^2}\left(\frac{A^4}{b^2}+\left(2\,\lambda +1\right)\left(A^2 -\lambda\right)-\lambda -2\right)+\frac{\gamma}{b}\left(A^2 +b^2\right)-\frac{A^2}{\lambda^2}\log\,b\Bigg.\Bigg]. \label{axialforcecase2}
\end{align}
Alternatively, we may fix $\mathcal{N}$ and $(\ref{axialforcecase2})$ can instead be solved for $\gamma=\gamma\,(\lambda)$.
In Fig. $\ref{fig2}$ we have shown the variation of $\mathcal{N}$ against $\lambda$ for three fixed values of $\gamma$, and $\gamma$ against $\lambda$ for three fixed values of $\mathcal{N}$. For the case $A=0.5$ considered, $\mathcal{N}$ has a maximum and a minimum if $\gamma>\gamma_{\text{min}}=8.46454$, and $\gamma$ has a maximum and a minimum if $\mathcal{N}>\mathcal{N}_{\text{min}}=33.2479$. At the respective thresholds $\gamma=\gamma_{\text{min}}$ and $\mathcal{N}=\mathcal{N}_{\text{min}}$, $\mathcal{N}$ and $\gamma$ have an inflection point at $\lambda_{\text{min}}=1.14282$. \\
As is the case for a solid cylinder \citep{FuST}, the condition for localised bifurcation corresponds to $\mathcal{N}$ (resp. $\gamma$) as a function of $\lambda$ attaining its extrema with $\gamma$ (resp. $\mathcal{N}$) fixed. For instance, we may deduce from $\partial \mathcal{N}/\partial \lambda=0$ the following expression for the critical surface tension $\gamma_{cr}$ at which a localised solution occurs
\begin{align}
\gamma_{cr}&=\frac{1}{b\,\lambda^{2}\left(A^2-1\right)}\left[\frac{4 \left(A^3 (\lambda -1)+A\right)^2 \log b}{A^2-1}-A^4\,\xi_{1}(\lambda) -A^2\,\xi_{2}(\lambda)-2 \left(\lambda ^3+2\right)\right], \label{bifconn}
\end{align}
where $\xi_{1}(\lambda)=2 \lambda ^5-4 \lambda ^4+2 \lambda ^3+2 \lambda ^2-3 \lambda +1$ and $\xi_{2}(\lambda)=4 \lambda ^4-4 \lambda ^3+7 \lambda -5$.
This expression has a minimum at $\lambda=\lambda_{\rm min}$, where $\lambda_{\rm min}$ depends on the tube's thickness.
As a form of validation, we take the limit of $(\ref{bifconn})$ as $A\rightarrow 0$ and obtain
\begin{align}
\gamma_{cr}&=\frac{2\left(2+\lambda^3\right)}{\lambda^{3/2}}. \label{gamcrsolid}
\end{align}
This is the well established condition for localised bifurcation in solid cylinders under axial stretching and surface tension given originally by \cite{taffetani}.

%%%%%%%%%%%%%%%%%%%%%%%%%%%%%%%%%%%%%%%%%%%%%%%%%%%%%%%%%%%%%%%%%%%%%%%%%%%%%%%%%%%%%%
%%%%%%%%%%%%%%%%%%%%%%%%%%%%%%%%%%%%%%%%%%%%%%%%%%%%%%%%%%%%%%%%%%%%%%%%%%%%%%%%%%%%%%

\subsubsection*{Case 3: Radially fixed outer boundary free of surface tension}
\begin{figure}[h!]
\centering
\begin{subfigure}[t]{0.486\textwidth}
\includegraphics[width=\linewidth, valign=t]{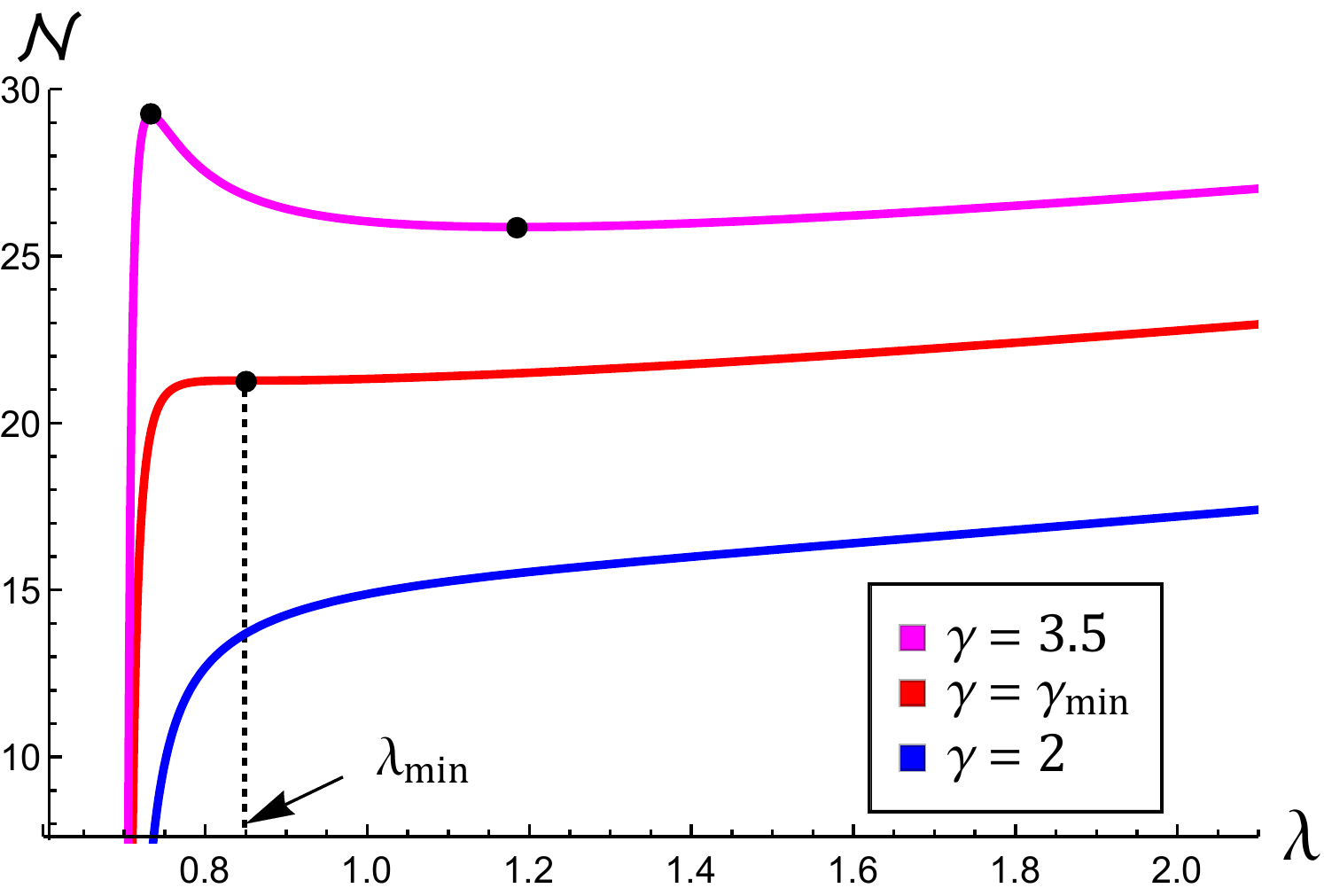}
\subcaption*{\textbf{(a)}}
\end{subfigure}\hfill
\begin{subfigure}[t]{0.493\textwidth}
\includegraphics[width=\linewidth, valign=t]{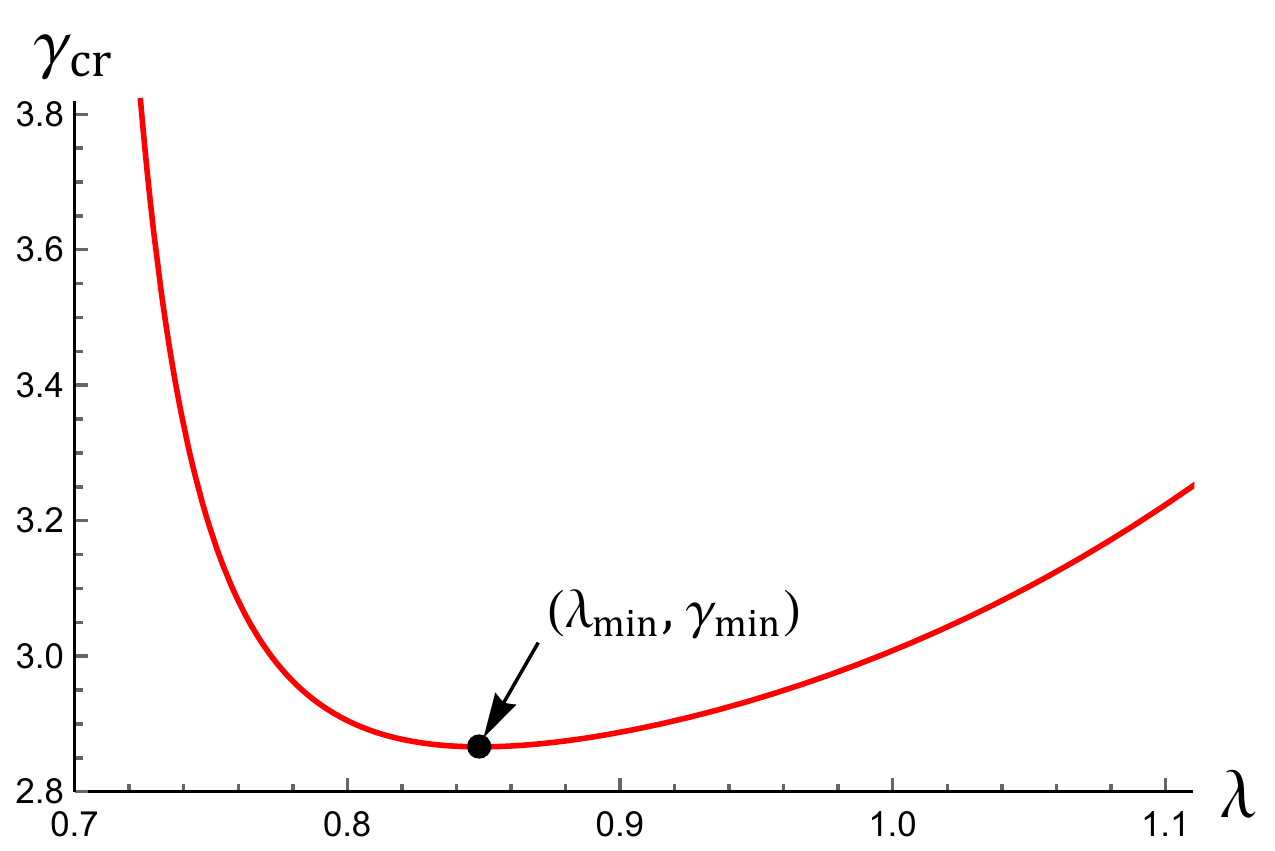}
\subcaption*{\textbf{(b)}}
\end{subfigure}
\caption{\textbf{(a)} Variation of $\mathcal{N}$ with respect to $\lambda$ for a tube of inner radius $A=0.55$ under several fixed surface tensions. Note that $\mathcal{N}$ is a monotonically increasing function of $\lambda$ for  $\gamma < \gamma_{\text{min}}$ and has a maximum and a minimum for $\gamma > \gamma_{\text{min}}$ where
$\gamma_{\text{min}}=2.86616$. The vertical order of the three curves and the legend parameter values are equivalent. \textbf{(b)} A blow up of the bifurcation condition \rr{gamcrcase3} about $\left(\lambda_{\text{min}},\,\gamma_{\text{min}}\right)$ where $\lambda_{\text{min}}=0.84881$ and $A=0.55$. Note that localised bifurcation cannot occur for $\gamma <\gamma_{\text{min}}$.}
\label{fig3}
\end{figure}

In case 3, the radial fixing of the outer boundary enforces the condition $b=B$, whilst the associated absence of surface tension requires that we set $\mathcal{E}_s^B=0$. With the aid of $(\ref{rR})$, the former condition is found to invoke the following expression for the finitely deformed inner radius $a$
\begin{align}
a=\sqrt{\lambda^{-1}\left(A^2 -1\right)+1}. \label{acase3}
\end{align}
Then, it follows that the primary solution $\phi_0$ in this case is given by
\begin{align}
\phi_0&=\frac{R^2 z}{2\,\lambda}+\frac{1}{2}\left(1-\frac{1}{\lambda}\right)z. \label{phi0case3}
\end{align}
Thus, as in case 2 previously, the primary deformation is determined solely by $\lambda$. From the equilibrium equation $\partial \mathcal{E}/\partial \lambda=0$, the following expression for $\mathcal{N}=\mathcal{N}(\lambda)$ is obtained
\begin{align}
\mathcal{N}&=\frac{\pi}{2\,\lambda^2}\Bigg[\Bigg.\frac{2\,\gamma\,\lambda^2}{a}\left(1+a^2\right)+\left(\lambda -1\right)\left(\frac{1}{a^2}+\lambda +1\right)-2\,A^2\left(\lambda^3 -1\right)+2\log\left(\frac{a}{A}\right)\Bigg.\Bigg]. \label{axialforcecase3}
\end{align}
We have shown in Fig. $\ref{fig3}$ (a) the variation of $\mathcal{N}$ against $\lambda$ for three fixed values of $\gamma$ and $A=0.55$, but the variation of $\gamma$ is not displayed for the sake of brevity. In this case the threshold value of $\gamma$ above which the variation of $\mathcal{N}$ is non-monotonic is $\gamma_{\text{min}}=2.86616$, and the threshold value of $\mathcal{N}$ above which the variation of $\gamma$ is non-monotonic is $\mathcal{N}_{\text{min}}=21.2744$. Then, as in case $2$, the condition for localised bifurcation is that $\mathcal{N}$ (resp. $\gamma$) as a function of $\lambda$ attains its extrema where $\gamma$ (resp. $\mathcal{N}$) is fixed. From $\partial \mathcal{N}/\partial \lambda=0$, an expression for $\gamma_{cr}$ is deduced
\begin{align}
\gamma_{cr}&=\frac{a}{\lambda^2\left(A^2-1\right)^2}\left[\left(2-2\,A^2\right)\,\lambda^4+\lambda+\lambda^2-4\,A^2\,\lambda-\lambda\,\log\left(\frac{a^4}{A^4}\right)+\frac{2+\lambda -\lambda^2}{a^4}\right]. \label{gamcrcase3}
\end{align}
This bifurcation condition is plotted in Fig. $\ref{fig3}$ (b) for the representative case $A=0.55$. It is observed that $\gamma_{cr}$ attains a minimum at $\lambda=\lambda_{\rm min}=0.84881$. Similar behaviour is observed in case 2 and also in the case of a solid cylinder. For the latter it is shown in \cite{FuST} that this minimum marks the transition from localised necking to localised bulging.

%%%%%%%%%%%%%%%%%%%%%%%%%%%%%%%%%%%%%%%%%%%%%%%%%%%%%%%%%%%%%%%%%%%%%%%%%%%%%%%%%%%%%%
%%%%%%%%%%%%%%%%%%%%%%%%%%%%%%%%%%%%%%%%%%%%%%%%%%%%%%%%%%%%%%%%%%%%%%%%%%%%%%%%%%%%%%

\section{Linear bifurcation analysis}
In the previous section we have presented for the three cases under consideration necessary conditions for localised bifurcation. Even if these are satisfied, we still need to ascertain whether localisation occurs before bifurcation into periodic modes. To this end, we solve in this section the eigenvalue problem governing infinitesimal perturbations of the primary solution and determine the dependence of the critical load on the axial wave number.

Consider a perturbation $\phi_1\left(R,z\right)$ of the finitely deformed state governed by $(\ref{phi0gen})$ or equivalent. On substituting the perturbed solution $\phi=\phi_0+\phi_1(R, z)$ into the equilibrium equation \rr{goveqn} and linearising in terms of $\phi_1(R, z)$, we obtain
\begin{align}
\mathcal{L}\left[\,\phi_{1}\,\right]+a_1(R)\,\phi_{1,RRzz} + a_2(R)\,\phi_{1,Rzz} + a_3(R)\,\phi_{1,zz} + a_4(R)\,\phi_{1,zzzz}&=0, \label{phi1GE}
\end{align}
where the operator $\mathcal{L}$ and the variable coefficients are given by
\begin{equation}
\begin{aligned}
\mathcal{L}\left[\,\phi\,\right]&=\phi_{,RRRR}-\frac{2}{R}\phi_{,RRR}+\frac{3}{R^2}\phi_{,RR}-\frac{3}{R^3}\phi_{,R},\,\,\,\,\,\,\,\,a_1(R)=\lambda^2+\frac{R^2}{\lambda \eta},\\[1em]
a_2(R)&=\frac{R\left(\eta -2\,R^2\right)}{\lambda\,\eta^2}-\frac{\lambda^2}{R},\,\,\,\,\,\,\,\,a_3(R)=\frac{2}{\lambda}\left(\frac{R^2}{\eta^2}-\frac{1}{R^2}\right)\,\,\,\,\,\,\,\,a_4(R)= \frac{\lambda R^2}{\eta}, \label{june3}
\end{aligned}
\end{equation}
with $\eta\equiv\eta(R)=R^2 - A^2 + a^2\,\lambda$. Equations \rr{phi1GE} -- \rr{june3} differ from equation (26) in Wang (2020). Agreement can only be achieved if we make the generally invalid substitution $a \to A/\sqrt{\lambda}$ in \rr{june3}, in which case the variable coefficients reduce to
\begin{align}
a_1(R)=\lambda^{-1}+\lambda^2,\,\,\,\,\,\,\,\,a_2(R)=-\frac{1+\lambda^3}{\lambda\,R},\,\,\,\,\,\,\,\,a_3(R)= 0,\,\,\,\,\,\,\,\,a_4(R)=\lambda. \label{june2}
\end{align}
The only exceptional case for which the above-mentioned substitution is valid is when the primary deformation is homogeneous. This may only be achieved when the outer radius tends to infinity or when $\lambda=1$ in cases 2 and 3 discussed previously, and a consequence of this is that incompressibility forces $ r=R/\sqrt{\lambda}$.

To validate our equations $(\ref{phi1GE})$ -- $(\ref{june3})$, we further make the  substitution $\phi_1=r f(r)\, {\rm e}^{\ii k z}$ and obtain a fourth-order differential equation for $f(r)$. We have verified that this equation is identical to the equation (53) of \cite{HO1979} when the latter is specialised to a neo-Hookean material.

We look for a non-trivial solution of the form
\begin{align}
\phi_1=   g\left(R\right)e^{ikz}, \label{phiincr}
\end{align}
where $k$ is the axial mode number and $g$ is a scalar function of $R$. On substituting $(\ref{phiincr})$ into $(\ref{goveqn})$, we obtain a fourth order ordinary differential equation (ODE) for $g$, which may be re-written as the following system of first order ODEs;
\begin{align}
\frac{d \bm{g}}{d R}&=\textsf{A}(R)\,\bm{g},\,\,\,\,\,\,\,\,
\textsf{A}=\begin{bmatrix}
0 & 1 & 0 & 0 \\
0 & 0 & 1 & 0 \\
0 & 0 & 0 & 1 \\
a_{41} & a_{42} & a_{43} & a_{44}
\end{bmatrix}, \label{goveq}
\end{align}
where $\bm{g}=\left[\,g,\,g',\,g'',\,g'''\,\right]^{\top}$ and the variable components of $\textsf{A}$ are given as follows
\begin{equation}
\begin{aligned}
a_{41}(R)&=k^2\left[\,\frac{2}{\lambda}\left(\frac{R^2}{\eta^2}-\frac{1}{R^2}\right)-\frac{k^2\lambda\,R^2}{\eta}\,\right],\,\,\,\,\,\,\,\,a_{42}(R)=k^2\left[\frac{R\left(\eta -2\,R^2\right)}{\lambda\,\eta^2}-\frac{\lambda^2}{R}\right]+\frac{3}{R^3},\\[1em]
a_{43}(R)&=k^2\left[\,\lambda^2+\frac{R^2}{\lambda \eta}\,\right]-\frac{3}{R^2},\,\,\,\,\,\,\,\,a_{44}(R)=\frac{2}{R}.
\end{aligned}
\end{equation}
On substituting $(\ref{phiincr})$ into $(\ref{BC1A})$ -- $(\ref{surflagrangianA})$ and $(\ref{BC2})$, we find that the boundary conditions on $R=A$ and $R=B$ in \textit{case 1} may be expressed as the following matrix equations
\begin{equation}
\begin{aligned}
\textsf{B}_1(A,\xi)\,\bm{g}&=\bm{0}, \\[1em]
\textsf{B}_2(B,\xi)\,\bm{g}&=\bm{0},
\end{aligned}\,\,\,\,\,\,\,\,\,\,\,\,\,\,\,\,\text{where}\,\,\,\,\,\,\,\,\,\,\,\,\,\,\,\,\begin{cases}
\textsf{B}_1(R,\xi)=\begin{bmatrix}
b_{11} & -1/R & 1 & 0 \\
b_{21}^{+} & b_{22} & -1/R & 1
\end{bmatrix}, \\[2em]
\textsf{B}_2(R,\xi)=\begin{bmatrix}
b_{11} & -1/R & 1 & 0 \\
b_{21}^{-} & b_{22} & -1/R & 1
\end{bmatrix},
\end{cases}
\end{equation}
with
\begin{equation}
\begin{aligned}
b_{11}(R,\,\xi)&=\frac{k^2 R^2}{\lambda\,\eta},\,\,\,\,\,\,\,\,b_{22}(R,\,\xi)=\frac{1}{R^2}-\frac{k^2}{\lambda\,\eta}\left[\,2\,R^2 + \lambda^3\,\eta\,\right],\\[1em]
b_{21}^{\pm}(R,\,\xi)&=k^2\,R\left[\,\frac{2\,R^4-\left(A^2-a^2\lambda\right)^2}{\lambda\,R^2\,\eta^2}\pm\frac{\gamma}{\lambda^{1/2}\eta^{3/2}}\left(k^2\,\eta - \lambda\right)\,\right].
\end{aligned}
\end{equation}
Note that $\xi$ is a dummy variable introduced for presentational purposes to represent the load parameter, for which there can be several choices. For cases 2 and 3, on substituting $(\ref{phiincr})$ into $(\ref{incphi})_1$, we deduce that satisfying zero incremental radial displacement on $R=A$ and $B$ (respectively) requires we enforce the corresponding constraints $g(A)=0$ and $g(B)=0$ in place of traction-free conditions. Indeed, the matrices $\textsf{B}_1$ and $\textsf{B}_2$ can then be modified accordingly. The linear system $\textsf{B}_1\left(A,\,\xi\right)\bm{g}=\bm{0}$ has two independent solutions, say $\bm{g}_{0}^{(1)}$ and $\bm{g}_{0}^{(2)}$. For instance, in case 1 we have
\begin{equation}
\begin{aligned}
\bm{g}_0^{(1)}&=\Big[\,1,\,0,\,-b_{11},\,-b_{11}/R-b^{+}_{21}\,\Big]^{\top}_{R=A},\,\,\,\,\,\,\,\,\bm{g}_0^{(2)}&=\Big[\,0,\,1,\,-1/R,\,1/R^2-b_{22}\,\Big]^{\top}_{R=A}. \label{IDa}
\end{aligned}
\end{equation}
We may then integrate forward $(\ref{goveq})$ from $R=A$ to $R=B$, using $(\ref{IDa})$ or equivalent as initial data for $\bm{g}$ at $R=A$. Two linearly independent solutions for $\bm{g}$, say $\bm{g}_1$ and $\bm{g}_2$ are obtained, and thus a general solution for $\bm{g}$ takes the form
\begin{align}
\bm{g}&=c_{1}\,\bm{g}_{1} + c_{2}\,\bm{g}_2=\textsf{M}\left(R,\,\xi\right)\bm{c}, \label{gensol}
\end{align}
where $\bm{c}=\left[\,c_{1},\,c_{2}\,\right]^{\top}$ is an arbitrary constant vector and $\textsf{M}\left(R,\,\xi\right)=\left[\,\bm{g}_{1},\,\bm{g}_2\,\right]$. By its construction, $(\ref{gensol})$ satisfies the boundary conditions on $R=A$, and it remains only to satisfy the corresponding conditions on $R=B$. On substituting $(\ref{gensol})$ into $\textsf{B}_2\left(B,\,\xi\right)\bm{g}=\bm{0}$, we obtain $\textsf{B}_2\,\textsf{M}\left(B,\,\xi\right)\bm{c}=\bm{0}$. Then, since $\bm{c}$ is arbitrary, the existence of a non-trivial solution to the eigenvalue problem is conditional on satisfying
\begin{align}
\text{det}\,\big[\,\textsf{B}_2\,\textsf{M}\left(B,\,\xi\right)\big] = 0. \label{NBC}
\end{align}
Thus, $(\ref{NBC})$ represents a numerical bifurcation condition which must be satisfied by $\gamma$, $\lambda$ and $k$. The bifurcation points are obtained by iterating on the load parameter $\xi$ until $(\ref{NBC})$ is satisfied. We may take either $\gamma$ or $\lambda$ as the load parameter.

The primary aim is to produce a numerical relationship between the load parameter $\xi$ and the axial mode number $k$. For a localised inhomogeneous solution to exist, we expect $\xi$ to take a physically plausible value at $k=0$ \citep{kirch,Iooss}. In such a case, we can then determine whether localisation is \textit{preferred} by the tube over periodic modes with $k \ne 0$. For instance, say we fix $\lambda$ and increase $\gamma$ monotonically from zero. Then, for localisation to be preferred we would expect curves in the $\left(k,\,\gamma\right)$ plane to have a minimum at $k=0$, and this is indeed the case for a solid cylinder. A minimum at a non zero value of $k$ indicates a preference towards periodic modes instead. We denote by $k_{cr}$ and $\gamma_{cr}$ the values of $k$ and $\gamma$ at this minimum.
%%%%%%%%%%%%%%%%%%%%%%%%%%%%%%%%%%%%%%%%%%%%%%%%%%%%%%%%%%%%%%%%%%%%%%%%%%%%%%%%%%%%%%
%%%%%%%%%%%%%%%%%%%%%%%%%%%%%%%%%%%%%%%%%%%%%%%%%%%%%%%%%%%%%%%%%%%%%%%%%%%%%%%%%%%%%%

\subsection*{Case 1: Traction-free curved boundaries under surface tension}
\begin{figure}[h!]
\centering
\begin{subfigure}[t]{0.49\textwidth}
\includegraphics[width=\linewidth, valign=t]{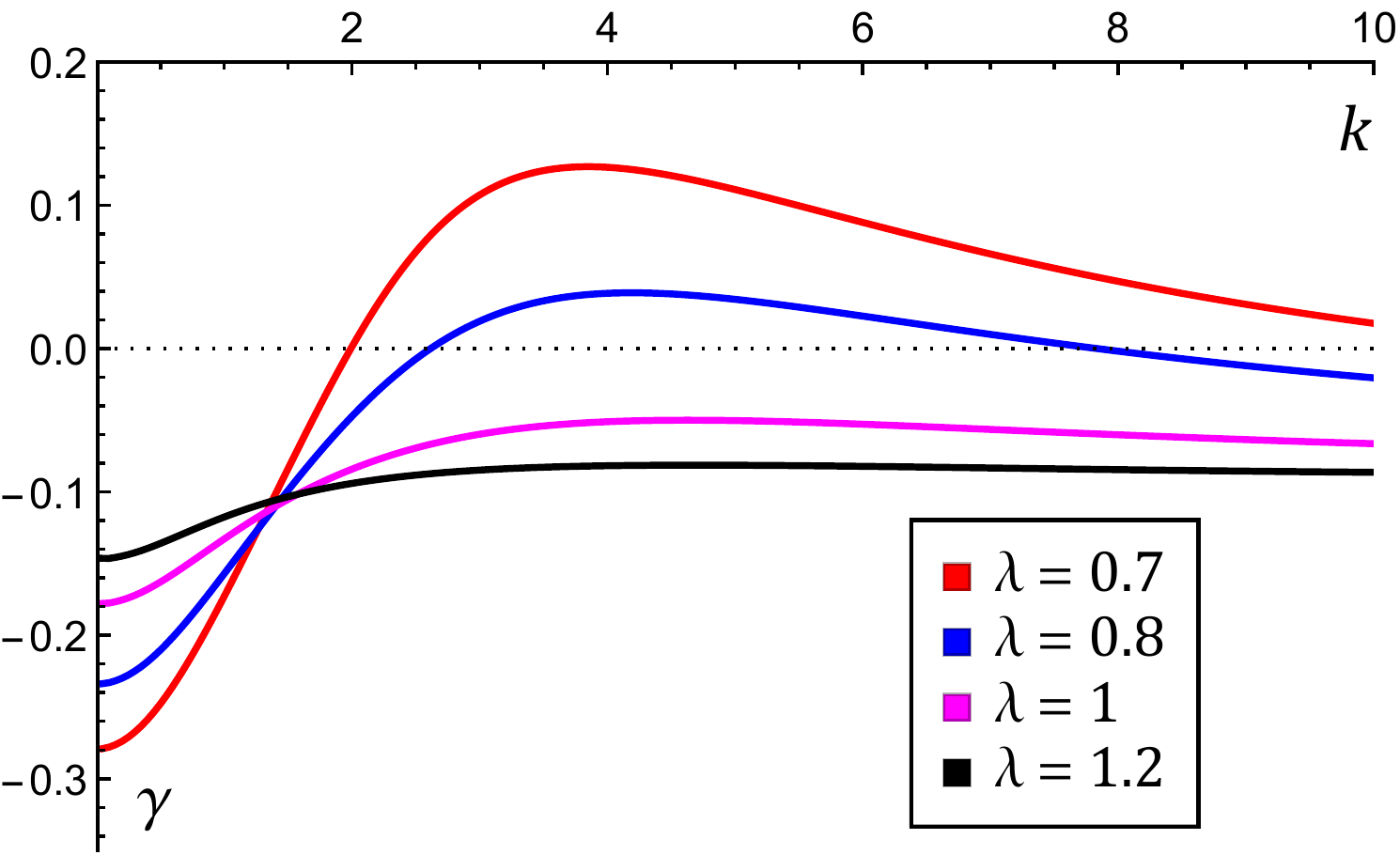}
\vspace{3mm}
\subcaption*{\textbf{(a)}}
\end{subfigure}\hfill
\begin{subfigure}[t]{0.475\textwidth}
\includegraphics[width=\linewidth, valign=t]{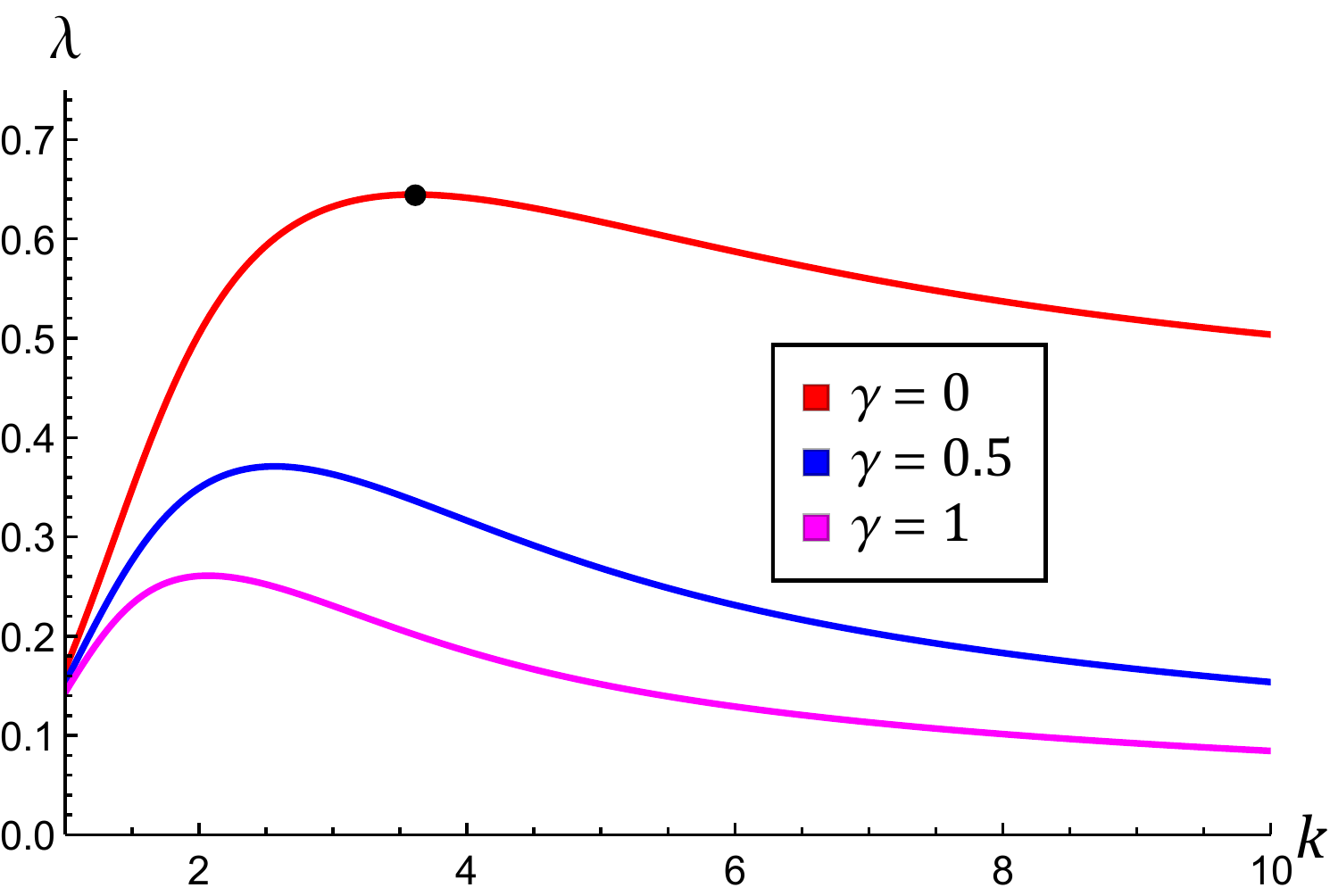}
\subcaption*{\textbf{(b)}}
\end{subfigure}
\caption{\textbf{(a)} The variation of $\gamma$ with respect to $k$ for $A=0.8$ and several fixed $\lambda$. \textbf{(b)} The variation of $\lambda$ with respect to $k$ for $A=0.5$ and several fixed $\gamma$. When $\gamma =0$ the preferred mode is $k=3.605$ occurring at the critical stretch $\lambda_{cr}=0.645$ (indicated by the black dot), which agrees with the result given by \citet{wi1955}. The vertical order of curves (for $k>2$, say, for \textbf{(a)}) and the legend parameter values are equivalent.}
\label{fig4}
\end{figure}

We begin by taking $\gamma$ as the load parameter, and plot the surface tension $\gamma$ against the mode number $k$ for several fixed $\lambda$ in Fig. $\ref{fig4}$ (a).
It is seen that the dependence of $\gamma$ on $k$ is very different from that in the case of solid cylinder. Fig. $\ref{fig4}$ (a) shows that no form of bifurcation can take place when the tube is stretched since $\gamma<0$ consistently. However, for sufficiently small fixed $\lambda<1$, whilst localised modes are still associated with negative surface tension values, bifurcation into non-zero periodic modes can be triggered at $\gamma=0$. To investigate this further, we consider the alternative loading condition whereby $\gamma$ is fixed and the tube is subjected to increasing compression. As shown in Fig. $\ref{fig4}$ (b), the tube gives preference towards periodic non-zero bifurcation modes rather than localised solutions in such a case. For larger fixed $\gamma$, the bifurcation curve descends, meaning that surface tension has a stabilising effect in the sense that it drastically decreases the critical stretch at which bifurcation can take place. When surface tension is sufficiently large, the bifurcation curve disappears completely and even bifurcation into a periodic mode becomes impossible.

We have numerically computed the relationship between $\gamma_{\rm cr}$ and $\lambda$ when $k=0$, and have verified that it is identical to \rr{jacobian} although both give negative values of $\gamma_{cr}$ which are physically unachievable.
%
% We also consider the variation of the critical deformed inner radius $a_{cr}$ at which localised bifurcation occurs against $\lambda$ for tubes of varying thickness. From Figure $\ref{lambdaacr}$ (b) we see that, when $a$ exceeds a certain value determined by $\lambda$ and $A$, surface tension becomes negligible. Clearly, for localised modes to exist in tubes under axial stretching and surface tension, we would expect the range of $a_{cr}$ shown in Figure $\ref{lambdaacr}$ (a) to coincide with the range of $a$ in which surface tension is prevalent. However, our findings are to the contrary of this.
%\begin{figure}[h!]
%\centering
%\begin{subfigure}[t]{0.37\textwidth}
%\includegraphics[width=\linewidth, valign=t]{acrgamplot}
%\subcaption*{\textbf{(a)}}
%\end{subfigure}\hfill
%\begin{subfigure}[t]{0.5575\textwidth}
%\includegraphics[width=\linewidth, valign=t]{Gammaplot}
%\subcaption*{\textbf{(b)}}
%\end{subfigure}
%\caption{\textbf{(a)} Localised bifurcation conditions showing the variation of $a_{cr}$ against $\lambda$ for various tube thickness's. \textbf{(b)} The variation of $\gamma$ (as given by $(\ref{gamcase1})$) against $a$ for various tube thickness's where $\lambda =1.2$.}
%\label{lambdaacr}
%\end{figure}

%%%%%%%%%%%%%%%%%%%%%%%%%%%%%%%%%%%%%%%%%%%%%%%%%%%%%%%%%%%%%%%%%%%%%%%%%%%%%%%%%%%%%%
%%%%%%%%%%%%%%%%%%%%%%%%%%%%%%%%%%%%%%%%%%%%%%%%%%%%%%%%%%%%%%%%%%%%%%%%%%%%%%%%%%%%%%

\subsection*{Case 2: Radially fixed inner boundary free of surface tension}
\begin{figure}[h!]
\centering
\begin{subfigure}[t]{0.486\textwidth}
\includegraphics[width=\linewidth, valign=t]{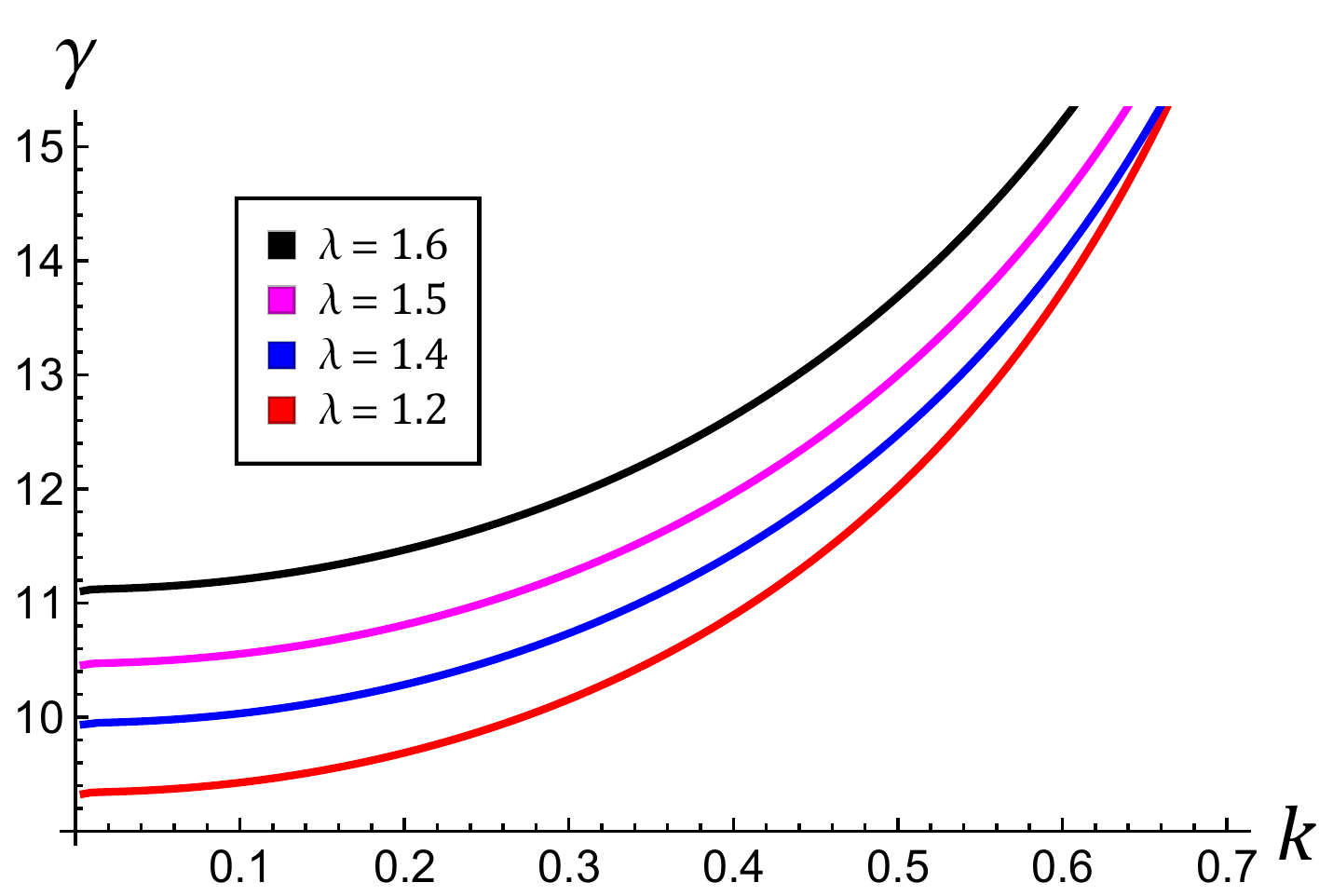}
\subcaption*{\textbf{(a)}}
\end{subfigure}\hfill
\begin{subfigure}[t]{0.495\textwidth}
\includegraphics[width=\linewidth, valign=t]{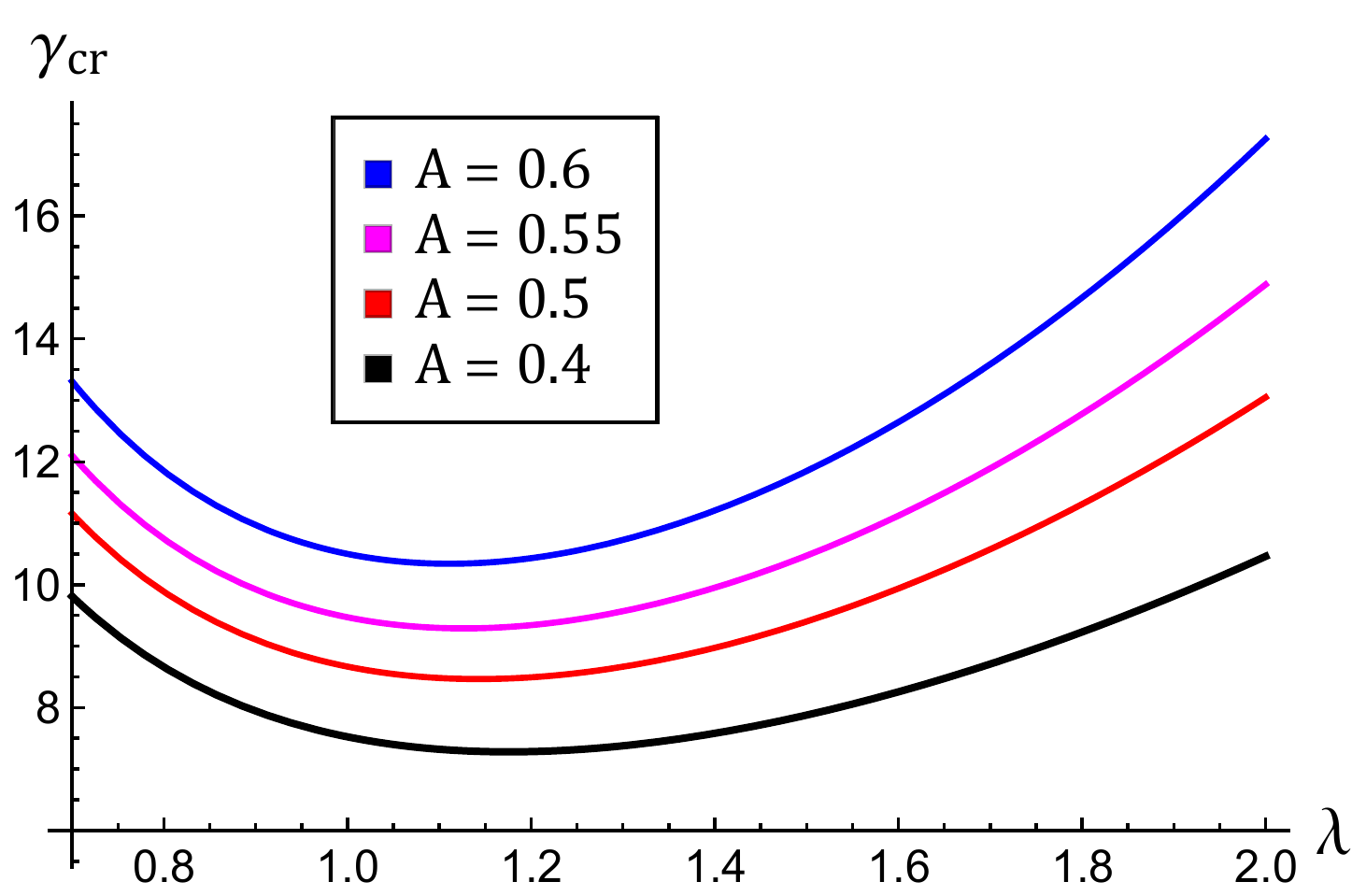}
\subcaption*{\textbf{(b)}}
\end{subfigure}
\begin{subfigure}[t]{0.475\textwidth}
\includegraphics[width=\linewidth, valign=t]{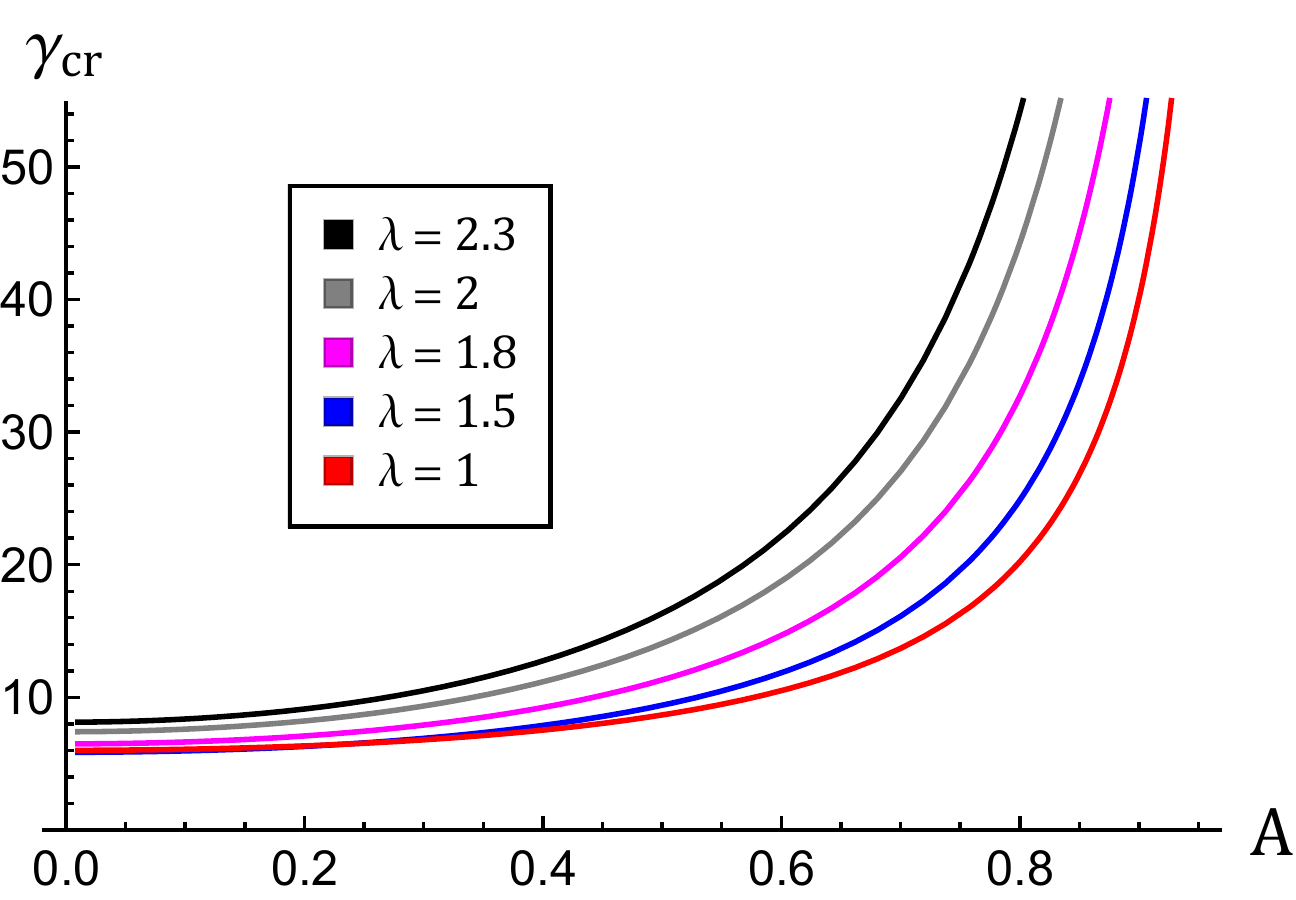}
\subcaption*{\textbf{(c)}}
\end{subfigure}\hfill
\begin{subfigure}[t]{0.485\textwidth}
\includegraphics[width=\linewidth, valign=t]{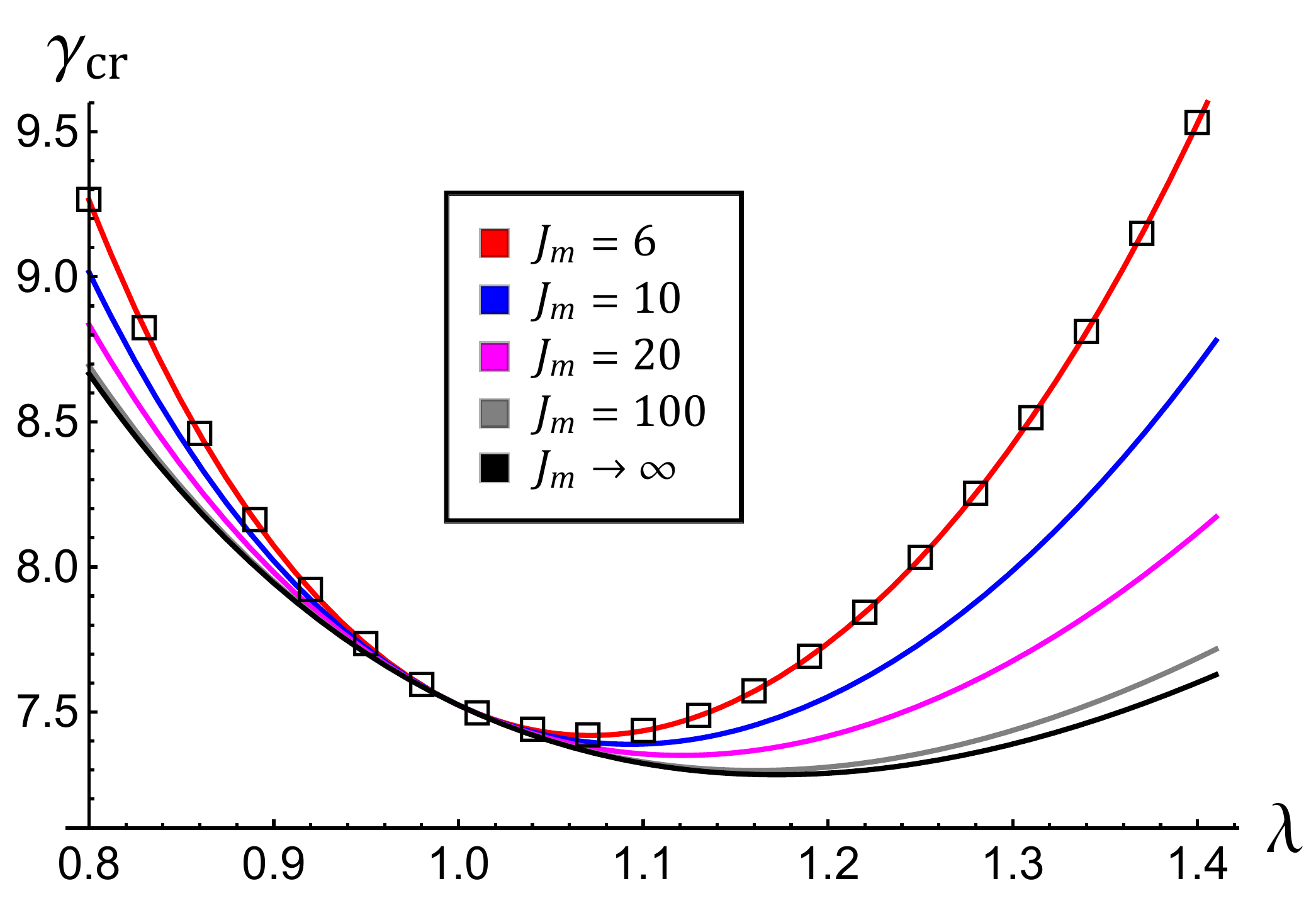}
\subcaption*{\textbf{(d)}}
\end{subfigure}
\caption{\textbf{(a)} The variation of $\gamma$ with respect to $k$ for $A=0.55$ and several fixed $\lambda$. In \textbf{(b)} and \textbf{(c)} we give conditions for bifurcation at $k=0$. For the neo-Hookean model, we present the variation of $\gamma_{cr}$ with respect to \textbf{(b)} $\lambda$ for several fixed $A$ and \textbf{(c)} $A$ for several fixed $\lambda$. In \textbf{(d)}, we present the Gent Model counterpart of the analytical condition $(\ref{bifconn})$. We plot $\gamma_{cr}$ against $\lambda$ for $A=0.4$ and several fixed $J_m$. The black squares give the corresponding numerical results obtained for $J_m=6$. The vertical order of curves and legend parameter values are equivalent.}
\label{fig5}
\end{figure}

Results in case 2 are in stark contrast to those presented previously for case 1. In Fig. $\ref{fig5}$ (a) we plot the load parameter $\gamma$ against $k$ for $A=0.55$ and several fixed $\lambda\geq 1$. Interestingly, we observe that localised modes are both possible and favourable since $k_{\rm cr}=0$ for all stretches considered. In Fig. $\ref{fig5}$ (b) and (c), the variation of the critical surface tension for localisation across different axial stretches and tube thickness's is considered.
\noindent We observe from Fig. $\ref{fig5}$\,(b) that $\gamma_{cr}$ as a function of $\lambda$ possesses a minimum for all tube thickness's considered. Such a property resonates with results obtained from the linear analysis of solid cylinders by \cite{FuST}, and there is a potential that the distinction between localised solutions either side of this minimum shown in the solid case may also occur here. However, such a conjecture must be investigated through a weakly non-linear analysis since a linear analysis gives no information on the nature of localised solutions in the near-critical regime. From Fig. $\ref{fig5}$\,(c), we determine that $\gamma_{cr}$ is an increasing function of $A$. Thus, greater tube thickness destabilises the tube towards localisation. We checked and verified that all numerical conditions in Fig. $\ref{fig5}$\,(b) and (c) are identical to the analytical counterpart $(\ref{bifconn})$.
%Validation of our efforts here can be sought from \cite{henann}, who conducted numerical simulations on internally fixed tubes under outer surface tension and no axial stretching. Their predictions for the critical surface tension at which localisation occurs are in excellent agreement with our corresponding stability threshold (red curve) in Figure $\ref{gamcrlamcase2}$ (b).

We also deduced localised bifurcation conditions for the Gent material model $(\ref{neohook})_2$ analytically using the variational approach employed in the previous section.
These conditions are presented in Fig. $\ref{fig5}$ (d). We observe that $\gamma_{cr}$ as a function of $\lambda$ possesses a minimum for materials of any extensibility, and $\gamma_{cr}$ increases as $J_m$ decreases for each fixed $\lambda\neq 1$. Thus, materials of lesser extensibility under fixed stretch may withstand higher levels of surface tension before instability ensues.

%%%%%%%%%%%%%%%%%%%%%%%%%%%%%%%%%%%%%%%%%%%%%%%%%%%%%%%%%%%%%%%%%%%%%%%%%%%%%%%%%%%%%%
%%%%%%%%%%%%%%%%%%%%%%%%%%%%%%%%%%%%%%%%%%%%%%%%%%%%%%%%%%%%%%%%%%%%%%%%%%%%%%%%%%%%%%

\subsection*{Case 3: Fixed outer surface free of surface tension}
\begin{figure}[h!]
\centering
\begin{subfigure}[t]{0.493\textwidth}
\includegraphics[width=\linewidth, valign=t]{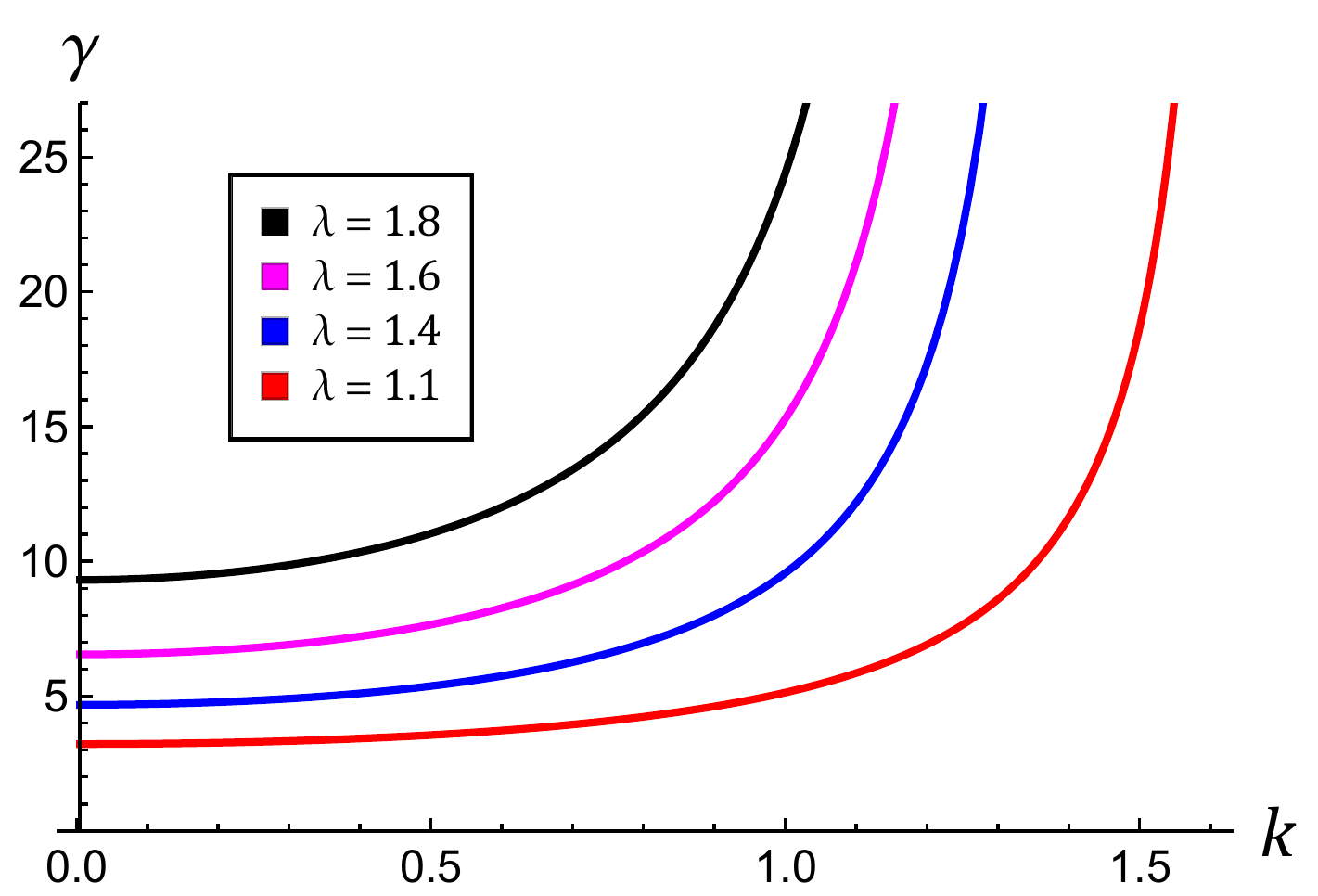}
\subcaption*{\textbf{(a)}}
\end{subfigure}\hfill
\begin{subfigure}[t]{0.493\textwidth}
\includegraphics[width=\linewidth, valign=t]{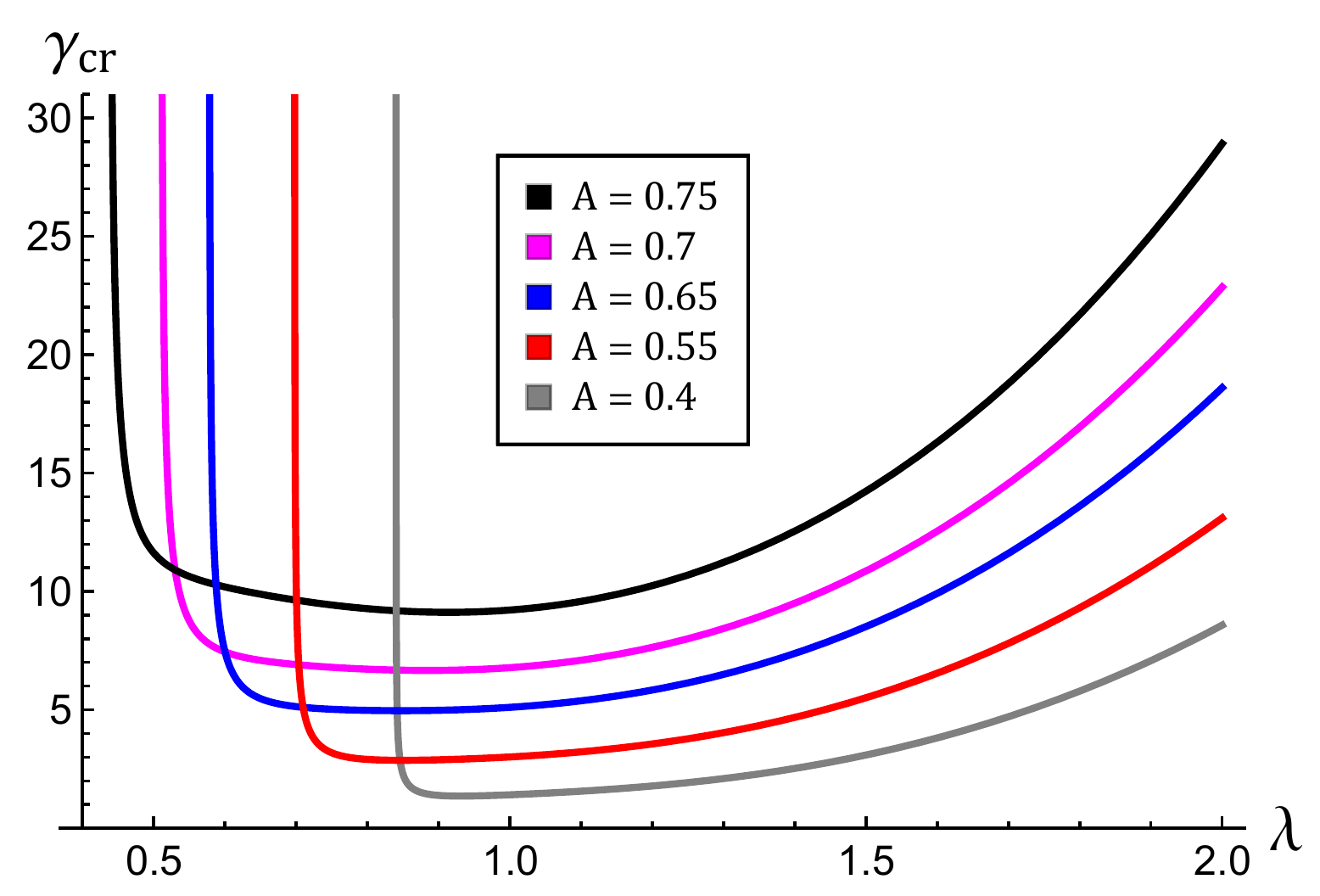}
\subcaption*{\textbf{(b)}}
\end{subfigure}
\begin{subfigure}[t]{0.5\textwidth}
\includegraphics[width=\linewidth, valign=t]{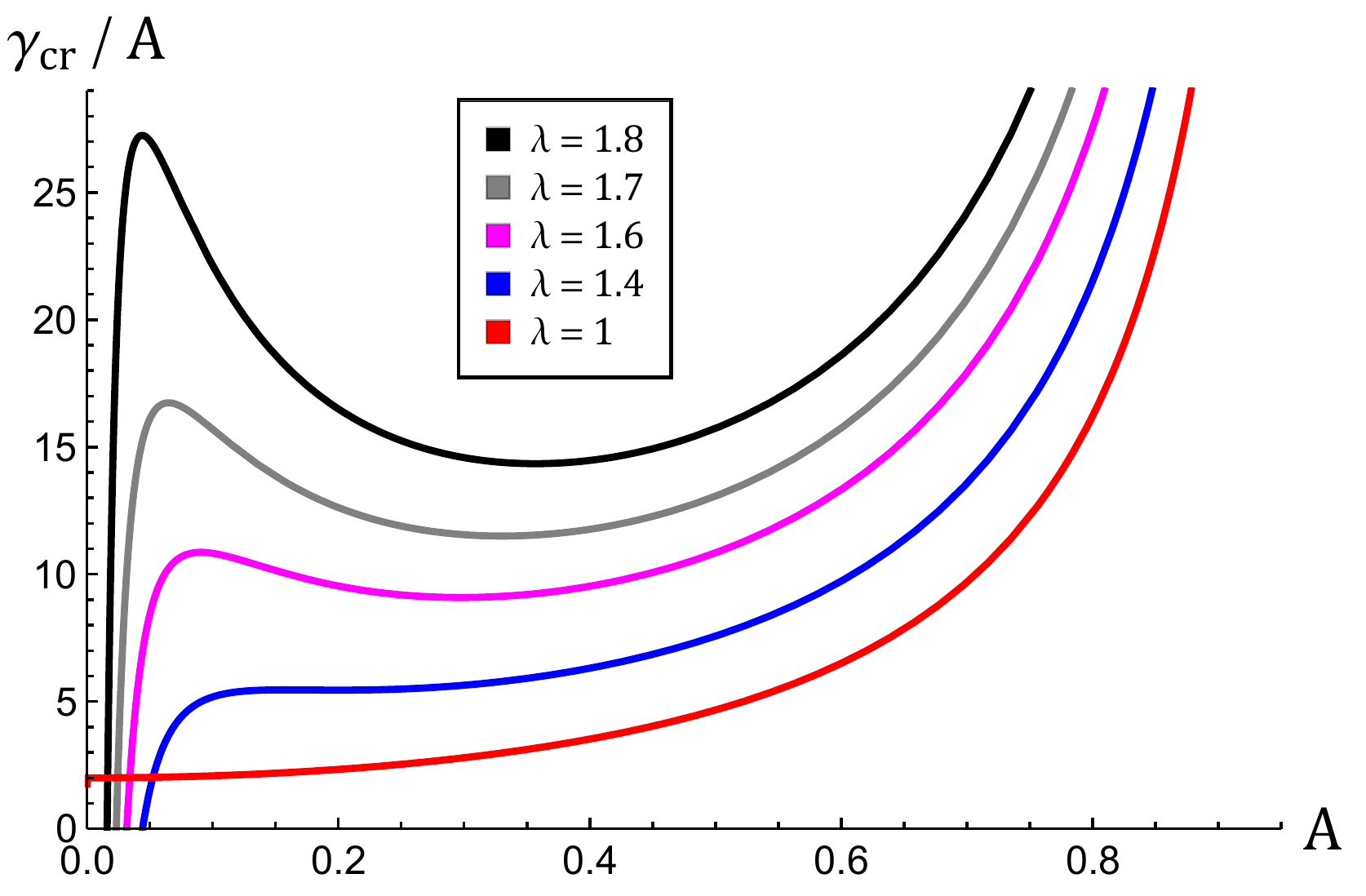}
\vspace{1mm}
\subcaption*{\textbf{(c)}}
\end{subfigure}\hfill
\begin{subfigure}[t]{0.49\textwidth}
\includegraphics[width=\linewidth, valign=t]{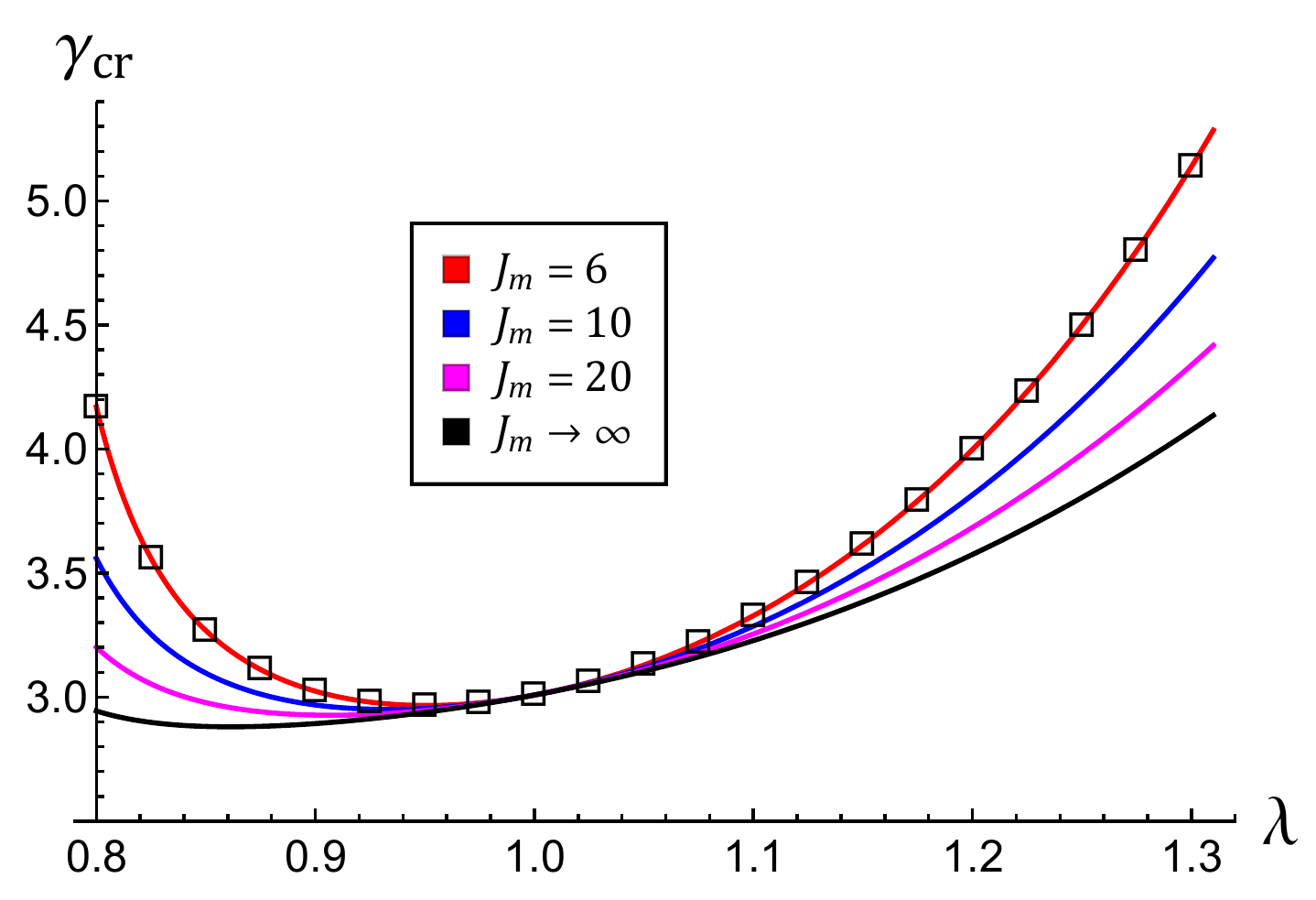}
\subcaption*{\textbf{(d)}}
\end{subfigure}
\caption{\textbf{(a)} The variation of $\gamma$ with respect to $k$ for $A=0.55$ and several fixed $\lambda$. In \textbf{(b)}, \textbf{(c)} and \textbf{(d)} we give conditions for bifurcation at $k=0$. For the neo-Hookean model, we present in \textbf{(b)} the variation of $\gamma_{cr}$ with respect to $\lambda$ for several fixed $A$, whilst in \textbf{(c)} we plot $\gamma_{cr}/A$ against $A$ for several fixed $\lambda$. In \textbf{(d)}, we present the Gent model counterpart of the analytical condition $(\ref{gamcrcase3})$. We plot $\gamma_{cr}$ against $\lambda$ for $A=0.55$ and several fixed $J_m$. The black squares give the corresponding numerical results obtained for $J_m=6$. The vertical order of curves and legend parameter values are equivalent.}
\label{fig6}
\end{figure}

In case 3, we again choose $\gamma$ as the load parameter and fix $\lambda$. We examine the variation of $\gamma$ against $k$ in Fig. $\ref{fig6}$ (a) for $A=0.55$ and several fixed $\lambda$. As in case $2$, the critical mode number is $k_{\rm cr}=0$ for all stretches considered. Therefore, a localised solution is not only possible but preferred over periodic modes.
We observe from Fig. $\ref{fig6}$ (b) that tubes can only admit a non-trivial localised solution up to a certain level of fixed compression. Indeed, as the axial stretch tends to some limiting value, $\gamma_{cr}$ is shown to diverge. It is also evident that the thicker the tube, the larger this limiting value becomes.
As in case $2$, the bifurcation curves possess minima; this has already been highlighted in Fig. $\ref{fig3}$ (b) for the representative case $A=0.55$. Thus a potential distinction between localised solutions either side of the critical stretch can also be pondered here. In Fig. $\ref{fig6}$ (c), we observe that for each fixed $\lambda>1$ considered, localisation is not possible beyond some critical tube thickness. For instance, for $\lambda=1.4$ (blue curve), the critical surface tension scaled by $A$ becomes negative below $A=0.0441709$. However, $\lambda=1$ (red curve) is an exception, and localisation can occur for any tube thickness. In fact, in the limit $A\rightarrow 0$, the case of a cylindrical cavity in an infinite solid is recovered, and we replicate the corresponding result $\gamma_{cr}/A \rightarrow 2$ that was given originally by \cite{xuan2016}. Indeed, the numerical bifurcation curves given in Fig. $\ref{fig6}$ (b) and (c) can also be obtained from our analytical condition $(\ref{gamcrcase3})$.

We also deduced localised bifurcation conditions for the Gent model $(\ref{neohook})_2$ analytically in Fig. $\ref{fig6}$ (d). As in case $2$, we observe that materials with lower extensibility limits are more resistant to localised modes since, for each fixed $\lambda$, $\gamma_{cr}$ increases as $J_m$ decreases.

%%%%%%%%%%%%%%%%%%%%%%%%%%%%%%%%%%%%%%%%%%%%%%%%%%%%%%%%%%%%%%%%%%%%%%%%%%%%%%%%%%%%%%
%%%%%%%%%%%%%%%%%%%%%%%%%%%%%%%%%%%%%%%%%%%%%%%%%%%%%%%%%%%%%%%%%%%%%%%%%%%%%%%%%%%%%%

\section{Comparison with FEM simulations}
We firstly facilitate a comparison with the FEM simulations of \cite{henann} in order to verify our analytical results for cases $2$ and $3$. In order to do so, we revisit our conditions analogous to $(\ref{bifconn})$ and $(\ref{gamcrcase3})$ for the Gent material model. As is done in the aforementioned simulations, we also assume that $\lambda =1$ throughout.

\begin{figure}[h!]
\centering
\begin{subfigure}[t]{0.492\textwidth}
\includegraphics[width=\linewidth, valign=t]{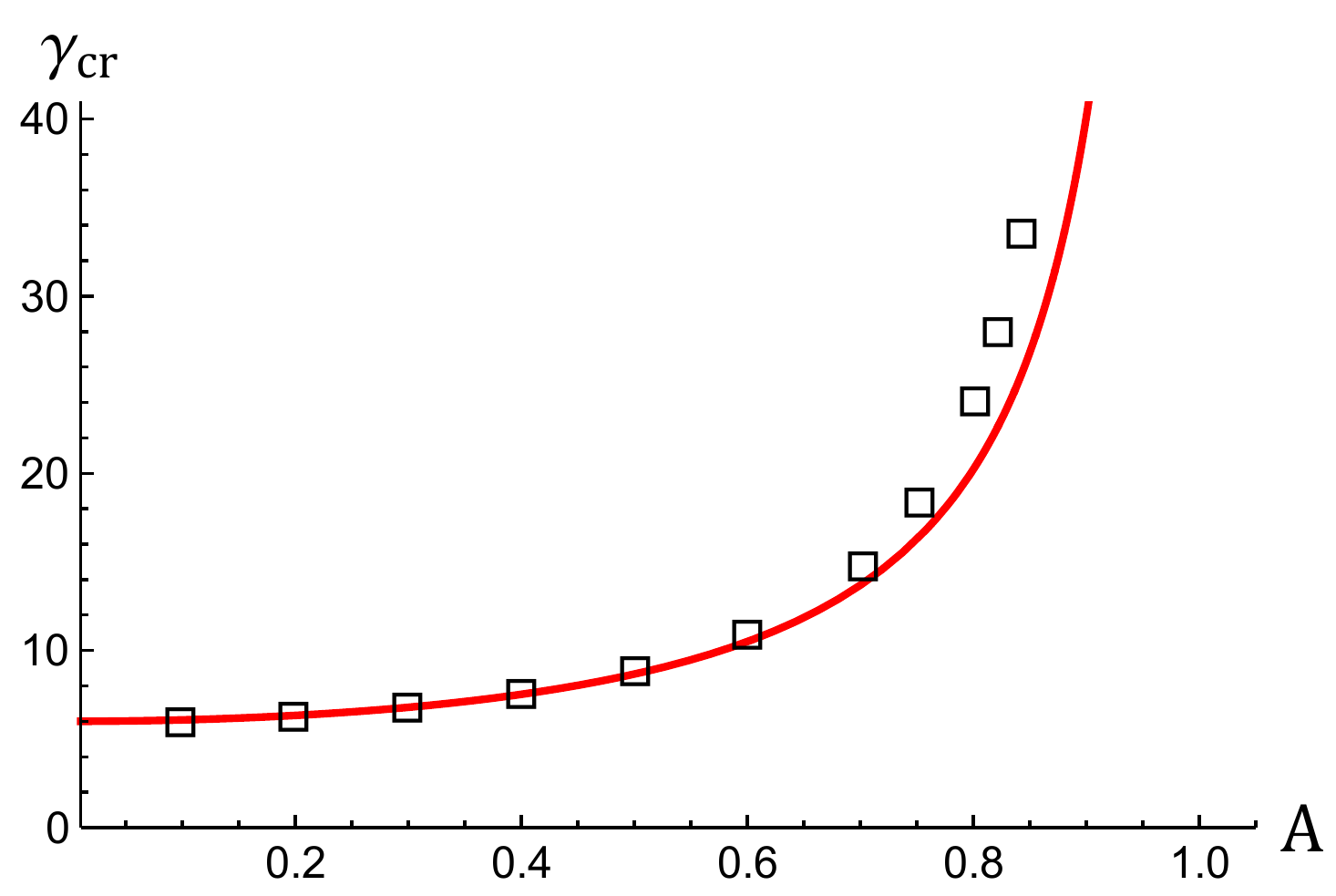}
\subcaption*{\textbf{(a)}}
\end{subfigure}\hfill
\begin{subfigure}[t]{0.5\textwidth}
\includegraphics[width=\linewidth, valign=t]{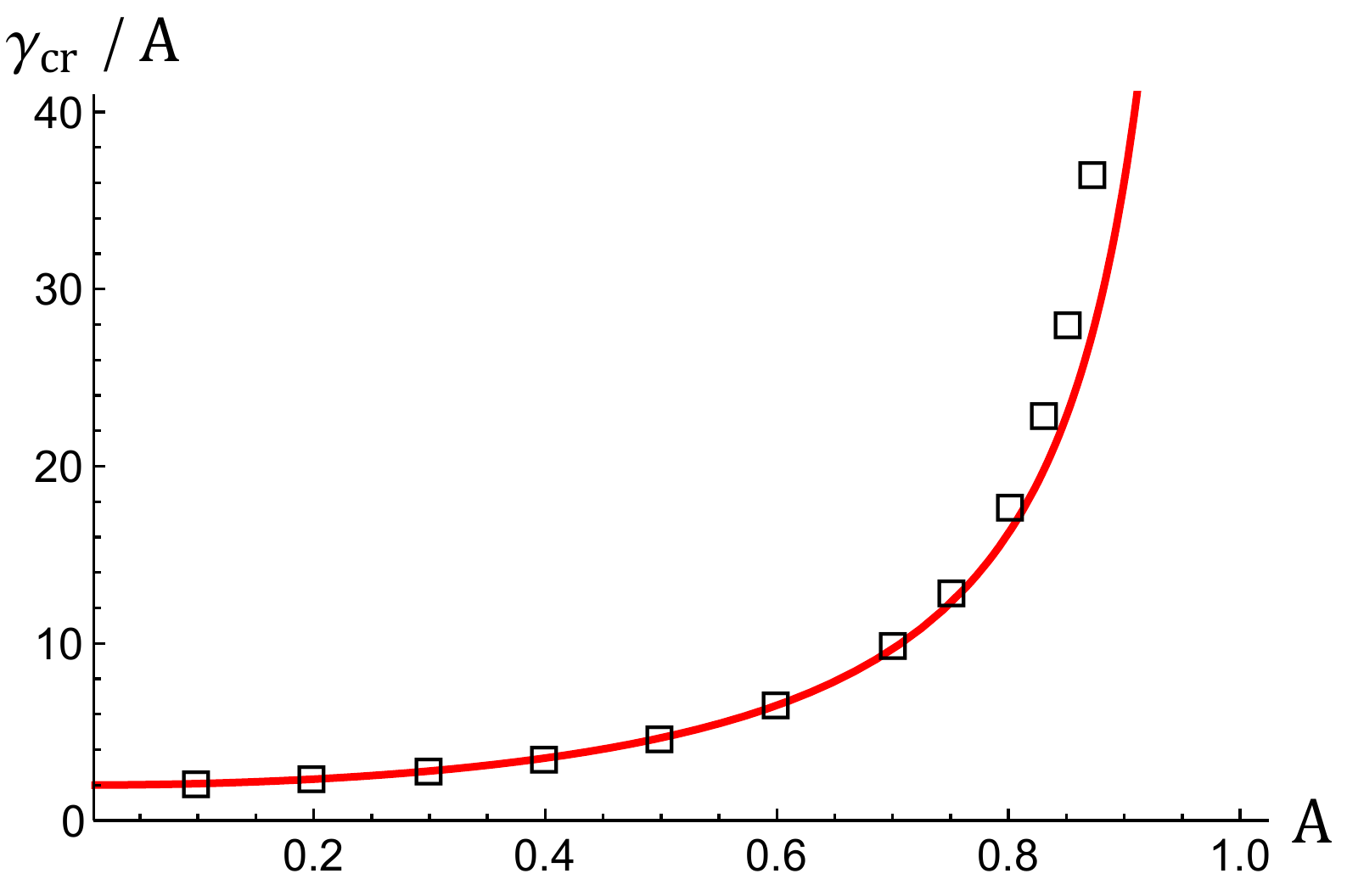}
\subcaption*{\textbf{(b)}}
\end{subfigure}
\caption{Localised bifurcation conditions for cases $2$ and $3$ respectively where the Gent material model is employed and $\lambda =1$. \textbf{(a)} The variation of $\gamma_{cr}$ as given by $(\ref{gentbifcon})_1$ against $A$ (red curve). Black squares give the corresponding FEM simulations in Fig. $4$ (b) of \cite{henann}. \textbf{(b)} The variation of $\gamma_{cr}/A$ as given by $(\ref{gentbifcon})_2$ against $A$ (red curve). Black squares give the corresponding FEM simulations in Fig. $4$ (c) of \cite{henann}.}
\label{fig7}
\end{figure}

It is noted that in both cases $2$ and $3$, the bifurcation condition is independent of the extensibility constant $J_m$ where $\lambda =1$. These conditions are given respectively as follows
\begin{align}
\gamma_{cr}&=\frac{2\left(3+A^2\right)}{1-A^2},\,\,\,\,\,\,\,\,\,\,\text{and}\,\,\,\,\,\,\,\,\frac{\gamma_{cr}}{A}=\frac{2\left(1+3\,A^2\right)}{1-A^2}. \label{gentbifcon}
\end{align}
We observe that in the limit $A\rightarrow 0$, $(\ref{gentbifcon})_2$ reduces to $\gamma_{cr}/A=2$, which is the localisation threshold for a cylindrical cavity inside an infinite solid given originally by \cite{xuan2016}. The conditions $(\ref{gentbifcon})_{1,\,2}$ are shown in Fig. $\ref{fig7}$ (a) and (b) respectively along with the corresponding FEM simulations of \cite{henann}.

To further validate our theoretical predictions when $\lambda \ne 1$, we have conducted additional numerical simulations in \cite{ab2013} by adapting the user subroutines of \citet{henann}. In our simulations we take $\mu=20 {\rm Pa}$, $L=10 {\rm mm}$, $A=0.10 {\rm mm}$, and $B=0.25 {\rm mm}$  so that the scaled value of $A$ is $0.4$.
Our simulations are conducted for the Gent material model with $J_m=100$ for which $\lambda_{\rm min}=1.161$ and $\gamma_{\rm min}=7.299$. We consider case 2 and focus on the scenario in which localisation/bifurcation is induced by increasing $\gamma$ gradually with the axial stretch fixed. We further assume that the total tube length is fixed during the entire process, that is both before and after bifurcation has taken place. This means that the {\it average axial stretch}, which is defined as the deformed length divided by the undeformed length, is fixed. For the bifurcation value of $\gamma$, we have excellent agreement between the simulation result and the theoretical prediction given by the counterpart of $(\ref{bifconn})$ for the Gent material considered.

%\begin{figure}[h!]
%\begin{center}
%\begin{tabular}{cc}
 %\includegraphics[scale=0.35]{fig11a} & \hspace{1cm}\includegraphics[scale=0.35]{fig11b} \\
 %(a) & (b)

%\end{tabular}
%\caption{Abaqus simulation results (solid lines) and theoretical predictions (squares) for the case when $\lambda$ is fixed to be $\lambda_{\rm min}$ and localization is induced by increasing $\gamma$. The dot represents the theoretical result given by the counterpart of \rr{gamcrcase2} for the Gent material.
%The two solid lines in (a) correspond to the axial stretches at the central section and at the two ends, respectively, computed according to $\lambda=(1-A^2)/(r^2-A^2)$ where $r$ is taken to be the radius of the outer surface at the appropriate cross-section. The total axial length is fixed throughout the entire process and  Gent material model with $J_m=100$ is used for which $\lambda_{\rm min}=1.161$ and $\gamma_{\rm min}=7.299$. For all $\gamma>\gamma_{\rm min}$ the deformation is always a kink-wave solution with the axial stretches in the two (almost) uniform sections satisfying Maxwell's equal area rule in the $F$ vs $\lambda$ diagram.}
%\label{fig11}
%\end{center}
%\end{figure}

\begin{figure}[h!]
\centering
\begin{subfigure}[t]{0.49\textwidth}
\includegraphics[width=\linewidth, valign=t]{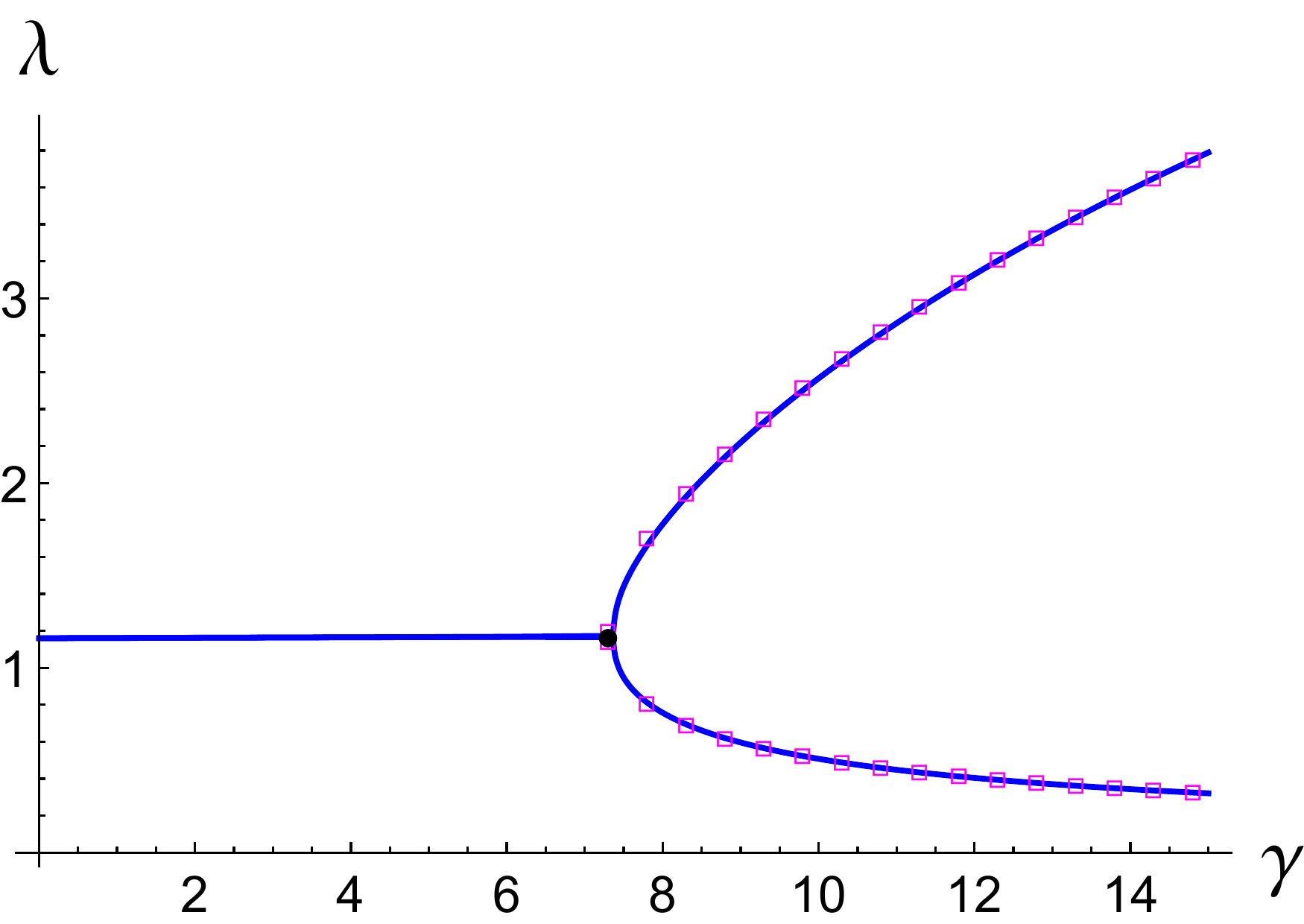}
\subcaption*{\textbf{(a)}}
\end{subfigure}\hfill
\begin{subfigure}[t]{0.49\textwidth}
\includegraphics[width=\linewidth, valign=t]{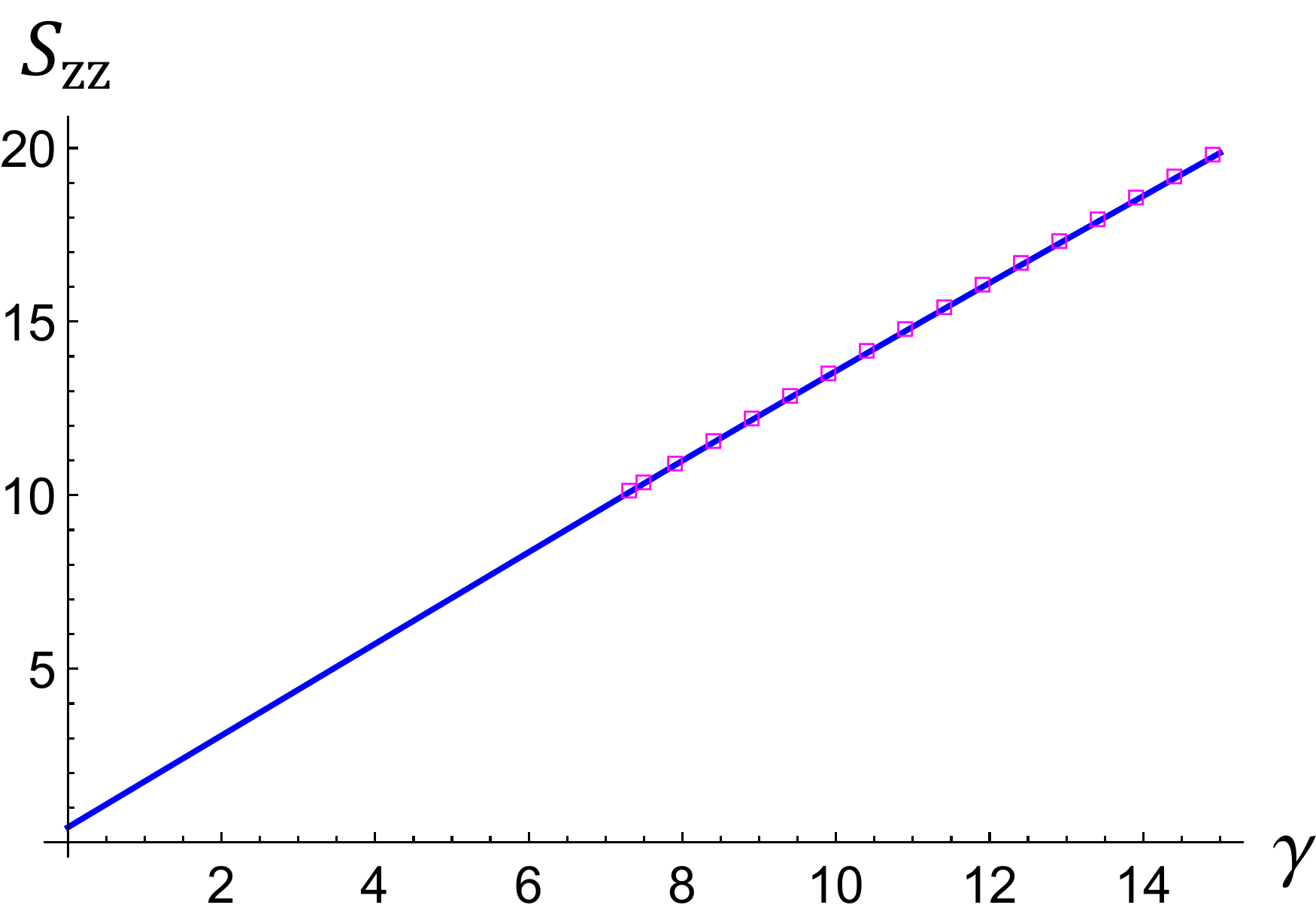}
\subcaption*{\textbf{(b)}}
\end{subfigure}
\caption{Abaqus simulation results (solid lines) and theoretical predictions (squares) for the case when $\lambda$ is fixed to be $\lambda_{\rm min}$ and localisation is induced by increasing $\gamma$. The single black dot represents the theoretical result given by the counterpart of \rr{bifconn} for the Gent material.
The lower and upper solid lines in (a) correspond to the axial stretches at the central (bulged) section and at the two (depressed) ends, respectively, computed according to $\lambda=(1-A^2)/(r^2-A^2)$ where $r$ is taken to be the radius of the outer surface at the appropriate cross-section. The total axial length is fixed throughout the entire process and  Gent material model with $J_m=100$ is used for which $\lambda_{\rm min}=1.161$ and $\gamma_{\rm min}=7.299$. For all $\gamma>\gamma_{\rm min}$ the deformation is always a kink-wave solution with the axial stretches in the two uniform sections satisfying Maxwell's equal area rule in the $S_{zz}$ vs $\lambda$ diagram.}
\label{fig8}
\end{figure}
\begin{figure}[h!]
\centering
\begin{subfigure}[t]{0.47\textwidth}
\includegraphics[width=\linewidth, valign=t]{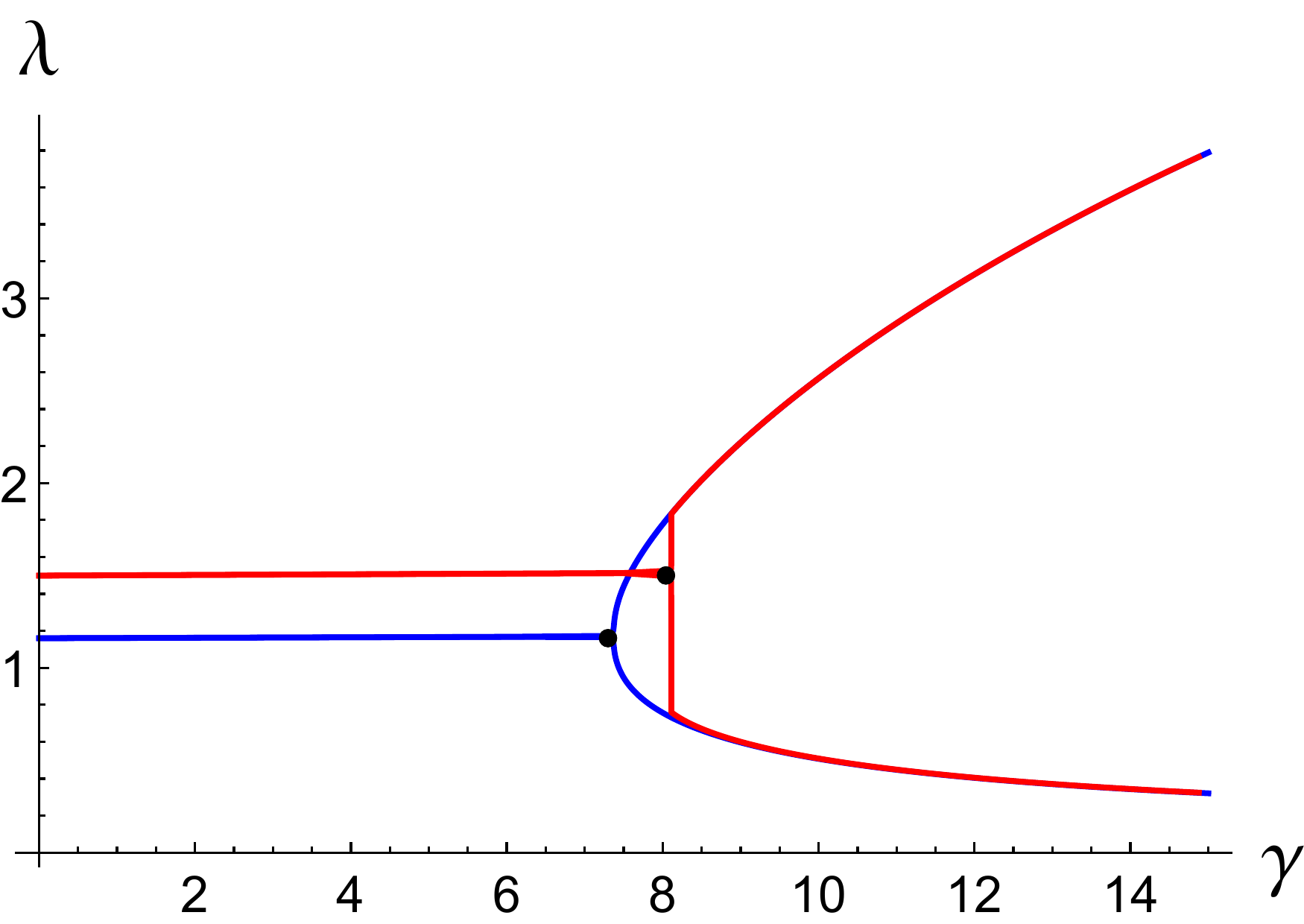}
\subcaption*{\textbf{(a)}}
\end{subfigure}\hfill
\begin{subfigure}[t]{0.47\textwidth}
\includegraphics[width=\linewidth, valign=t]{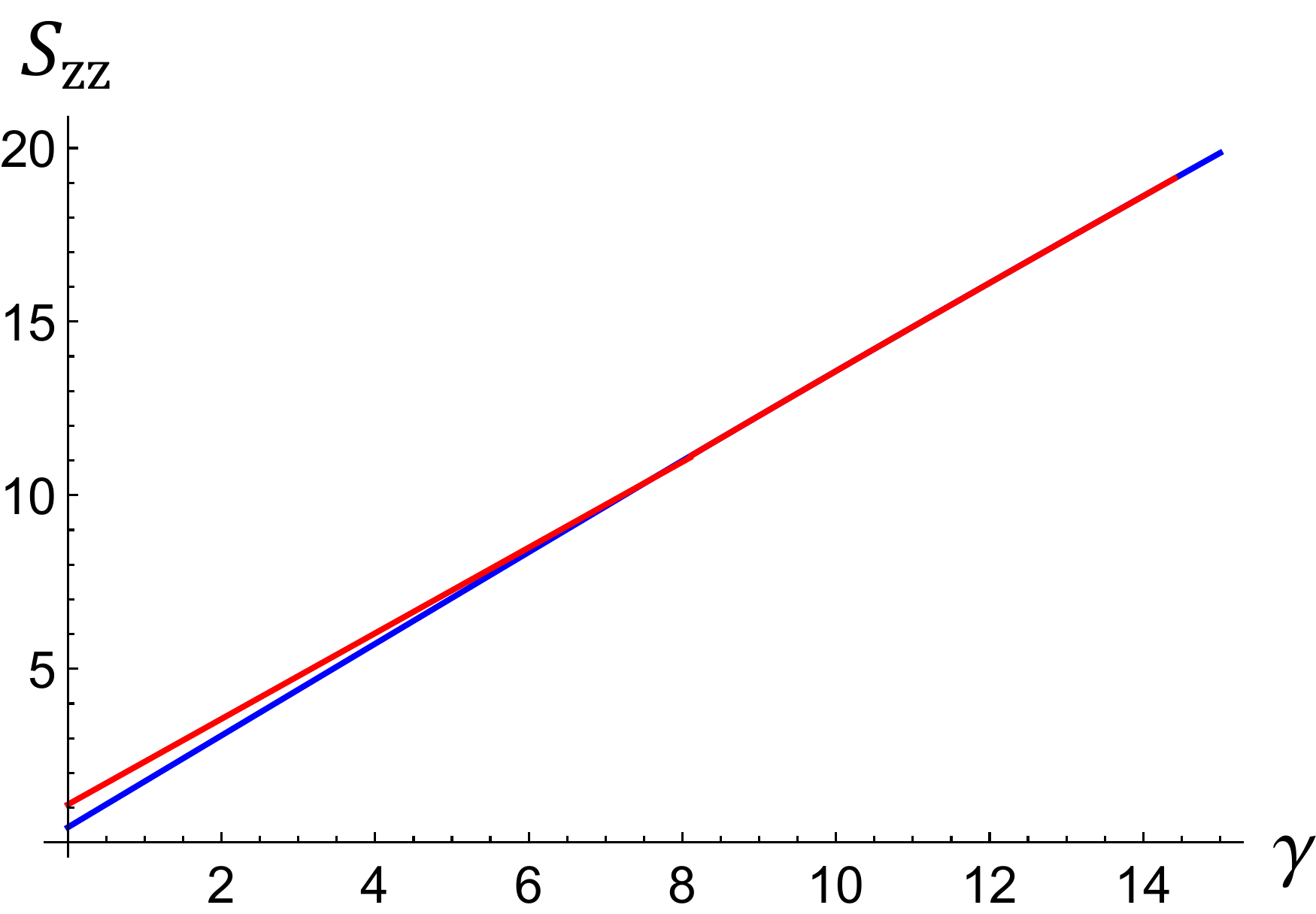}
\subcaption*{\textbf{(b)}}
\end{subfigure}
\caption{Abaqus simulation results when $\lambda$ is fixed to be $\lambda_{\rm min}$ (blue lines) and $1.5$ (red lines), respectively, and Gent material model with $J_m=100$ is employed. The two black dots represent the theoretical results given by the counterpart of \rr{bifconn} for the Gent material. It is noted that although when $\lambda =1.5$ bifurcation takes place later, the associated axial stretches in the center and at the two ends will jump to join the curves corresponding to $\lambda =\lambda_{\rm min}$. This means that no matter what value $\lambda$ takes, the tube always adopts the same kink-wave state that is determined by the value of $\gamma$ and the corresponding $S_{zz}$ vs $\lambda$ diagram.}
\label{fig9}
\end{figure}
\begin{figure}[h]
\begin{center}
%\begin{tabular}{cc}
% \includegraphics[width=.618\textwidth]{fig10a}
  \includegraphics[width=.9\textwidth]{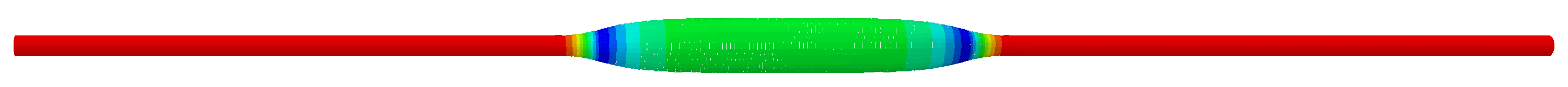}
 \end{center}
%\textbf{(b)} & \includegraphics[width=.8\textwidth]{fig10b}\\ \\
%\end{tabular}
\caption{Profile of the tube when the average axial stretch is fixed to be $\lambda_{\rm min}$ and $\gamma$ is increased to $9$. The values of $\lambda$ in the two \lq\lq phases" are determined by \rr{june11} and the proportion of each phase is determined by the average axial stretch imposed.}
\label{fig10}

\end{figure}

Although we have only focused on a linear analysis, our expressions for the primary deformation can in fact be used to predict the fully developed \lq\lq two-phase" deformation that the tube will adopt after going through the initial bifurcation. Thus, more comparisons/validations can be made in addition to the comparison made above for the bifurcation value of $\gamma$. Extending the observations made by  \cite{xuan2017} and \cite{FuST} for the case of a solid cylinder, we may summarize the anticipated bifurcation behaviour as follows. When $\lambda=\lambda_{\rm min}$, bifurcation will take place when $\gamma$ reaches the critical value $\gamma_{\rm min}$. For each $\gamma>\gamma_{\rm min}$, the curve of $\mathcal{N}$ against $\lambda$ has a maximum and a minimum, and by applying the equal area rule we may determine two values $\lambda_{\rm S}$ and $\lambda_{\rm L}$ such that
$\lambda_{\rm S}< \lambda_{\rm L}$, and
\be S_{zz}(\lambda_{\rm S})=S_{zz}(\lambda_{\rm L}), \;\;\;\; \int^{\lambda_{\rm L}}_{\lambda_{\rm S}} S_{zz} d \lambda =S_{zz}(\lambda_{\rm S}) (\lambda_{\rm L}-\lambda_{\rm S}), \la{june11} \en
where $S_{zz}$ denotes $\mathcal{N}$ scaled by the cross-sectional area $\pi (1-A)^2$. Note that $\lambda_{\rm S}$ and $\lambda_{\rm L}$ are in fact functions of $\gamma$. These functions are determined numerically on Mathematica for the Gent material model considered.  According to \cite{FuST}, as soon as $\gamma$ is increased beyond $\gamma_{\rm min}$, the deformation will be a static kink wave consisting of a bulged section with axial stretch $\lambda_{\rm S}$ and a depressed section with axial stretch $\lambda_{\rm L}$, the two sections being joined by a sharp but smooth transition region (similar to the coexistence of two-phases in one-dimensional phase transitions). The proportion of the bulged section is determined by the specified total length (or equivalently the average stretch). Furthermore, if $\lambda$ is fixed at a value other than $\lambda_{\rm min}$, bifurcation/localisation will take place at the value of $\gamma$ determined by the bifurcation condition, but as soon as $\gamma$ is increased above its bifurcation value, the tube will jump to the same kink wave configuration corresponding to $\lambda=\lambda_{\rm min}$ although the proportion of the bulged section will be different since the length is now fixed at a different value. These predictions are fully confirmed by our numerical simulations. In Fig. \ref{fig8} we show the perfect agreement between the simulation results and our theoretical results for the case when $\lambda=\lambda_{\rm min}$, whereas in Fig. \ref{fig9} we confirm the above-mentioned jump behaviour for a typical value of $\lambda=1.5$. Finally in Fig. \ref{fig10} we display a typical \lq\lq two phase" configuration of the tube when the average axial stretch is fixed to be $\lambda_{\rm min}$ and $\gamma$ is increased to $9$. All our numerical results have been obtained by
adopting the geometrical imperfection recommended by \citet{henann}, namely that the wall thickness is reduced linearly from both ends of the tube towards the middle section ($Z=0$) where the maximum reduction imposed is $0.004\%$.

%Note that the bulged section may appear in the middle or near the two ends depending on how we introduce the tiny imperfection in our simulations. We adopt the geometrical imperfection recommended by \citet{henann}, and Fig. \ref{fig10} (a) and (b) correspond to a localised narrowing and widening around the middle section, respectively.

%\begin{figure}[h]
%\begin{center}
%\begin{tabular}{cc}
% \includegraphics[width=.4\textwidth]{fig12a} & \hspace{1cm}\includegraphics[width=.4\textwidth]{fig12b} \\
 %(a) & (b)

%\end{tabular}
%\caption{Abaqus simulation results when $\lambda$ is fixed to be $\lambda_{\rm min}$ and $1.5$, respectively, and Gent material model with $J_m=100$ is employed. The dots represent the theoretical results given by the counterpart of \rr{gamcrcase2} for the Gent material. It is noted that although when $\lambda =1.5$ bifurcation takes place later, the associated axial stretches in the center and at the two ends will jump to join the curves corresponding to $\lambda =\lambda_{\rm min}$. This means that no matter what value $\lambda$ takes, the tube always adopts the same kink-wave state that is determined by the value of $\gamma$ and the corresponding $F$ vs $\lambda$ diagram.}
%\label{fig12}
%\end{center}
%\end{figure}

\section{Conclusion}
The objective of this study was two-fold. Firstly, determine a physical interpretation of localised bifurcation in cylindrical tubes under three separate constraints through analytical means. In case $1$, the inner and outer surfaces of the tube were traction-free and under surface tension, whilst in cases $2$ and $3$ the inner and outer surfaces respectively were fixed to prevent radial displacement and surface tension. Secondly, establish whether localisation is indeed possible and favoured in any of these 3 scenarios through a linear bifurcation analysis. We discovered that a condition for localised bifurcation can indeed be deduced analytically in each scenario and under any loading condition by applying the variational formulation in section $2$. For case $1$, we stated that localised bifurcation occurs where the Jacobian of the vector function $(\gamma,\,\mathcal{N})$ vanishes. In cases $2$ and $3$, localisation was found to occur when the resultant axial force as a function of the axial stretch attains its extrema for fixed surface tension, or when the surface tension as a function of the axial stretch attains it extrema for fixed axial force. Analytical bifurcation conditions were provided for each scenario. Of course, we recognised that these conditions are only valid provided that bifurcation into a localised solution is possible. To determine the existence of localisation, we conducted a linear analysis for all three cases. For case $1$ our analysis demonstrated that, when fixing the axial stretch $\lambda$ and taking the surface tension $\gamma$ as the load parameter, $\gamma\leq 0$ at $k=0$ for all $\lambda$ considered. This suggests that bifurcation into localised solutions is not possible in this case. Instead bifurcation into periodic modes with a preferred non-zero wave number is possible when the axial compression is sufficiently large and surface tension has a stabilizing effect in this respect. In contrast, localised bifurcation was shown to be both possible and favourable in cases $2$ and $3$. Our linear analysis showed that, where $\gamma$ is increased and $\lambda$ is fixed, the critical surface tension for a localised mode $\gamma_{cr}$ as a function of $\lambda$ possesses a minimum in both scenarios. It can then be expected, based on the weakly nonlinear analysis of \cite{FuST} for a solid cylinder, that a similar distinction, necking or bulging, between localised solutions either side of this minimum will also occur. We verified that the numerical bifurcation curves showing the variation of $\gamma_{cr}$ on $\lambda$ and tube thickness were in perfect agreement with our analytical conditions. Further validation comes from the excellent agreement of our results with the corresponding numerical simulation results.

We conclude by noting that the linear bifurcation condition derived in this paper is only a necessary condition for localisation to occur; whether such a bifurcation can really occur or not, whether the bifurcation is sensitive to imperfections, and whether the initial bifurcated configuration is a bulge or a depression can only be settled by a nonlinear analysis. Such a weakly non-linear analysis will be presented in a separate paper.

\subsection*{Acknowledgements}
The authors thank Dr. Lishuai Jin of Harvard University (now at University of Pennsylvania) for his help with the Abaqus simulations reported in this paper. The first author (DE) also acknowledges Keele University for supporting his PhD studies through a Faculty Studentship.
%Finally, we constructed a concrete argument which settled the discrepancy between the efforts shown here and those of \cite{liuwang}. By proving that, unlike in the aforementioned paper, our formulation reduces to that presented by \cite{HO1979}, we demonstrated that Wang's boundary value problem is in error, and that our paper is indeed valid.

\bibliographystyle{elsarticle-harv}
\bibliography{refer}

\begin{thebibliography}{43}
\expandafter\ifx\csname natexlab\endcsname\relax\def\natexlab#1{#1}\fi
\providecommand{\url}[1]{\texttt{#1}}
\providecommand{\href}[2]{#2}
\providecommand{\path}[1]{#1}
\providecommand{\DOIprefix}{doi:}
\providecommand{\ArXivprefix}{arXiv:}
\providecommand{\URLprefix}{URL: }
\providecommand{\Pubmedprefix}{pmid:}
\providecommand{\doi}[1]{\href{http://dx.doi.org/#1}{\path{#1}}}
\providecommand{\Pubmed}[1]{\href{pmid:#1}{\path{#1}}}
\providecommand{\bibinfo}[2]{#2}
\ifx\xfnm\relax \def\xfnm[#1]{\unskip,\space#1}\fi
%Type = Manual
\bibitem[{{Abaqus}(2013)}]{ab2013}
\bibinfo{author}{{Abaqus}}, \bibinfo{year}{2013}.
\newblock \bibinfo{title}{ABAQUS Analysis Users Manual, version 6.13}.
\newblock \bibinfo{note}{Dassault Systems, Providence, RI, USA}.
%Type = Article
\bibitem[{Alhayani et~al.(2014)Alhayani, Rodr{\'\i}guez and
  Merodio}]{alhayani2014competition}
\bibinfo{author}{Alhayani, A.}, \bibinfo{author}{Rodr{\'\i}guez, J.},
  \bibinfo{author}{Merodio, J.}, \bibinfo{year}{2014}.
\newblock \bibinfo{title}{Competition between radial expansion and axial
  propagation in bulging of inflated cylinders with application to aneurysms
  propagation in arterial wall tissue}.
\newblock \bibinfo{journal}{Int. J. Eng. Sci.} \bibinfo{volume}{85},
  \bibinfo{pages}{74--89}.
%Type = Article
\bibitem[{Barriere et~al.(1996)Barriere, Sekimoto and Leibler}]{bs1996}
\bibinfo{author}{Barriere, B.}, \bibinfo{author}{Sekimoto, K.},
  \bibinfo{author}{Leibler, L.}, \bibinfo{year}{1996}.
\newblock \bibinfo{title}{Peristaltic instability of cylindrical gels}.
\newblock \bibinfo{journal}{J. Chem. Phys.} \bibinfo{volume}{105},
  \bibinfo{pages}{1735--1738}.
%Type = Article
\bibitem[{Bico et~al.(2018)Bico, Reyssat and Roman}]{bico2018}
\bibinfo{author}{Bico, J.}, \bibinfo{author}{Reyssat, {\'E}.},
  \bibinfo{author}{Roman, B.}, \bibinfo{year}{2018}.
\newblock \bibinfo{title}{Elastocapillarity: When surface tension deforms
  elastic solids}.
\newblock \bibinfo{journal}{Annu. Rev. Fluid Mech.} \bibinfo{volume}{50},
  \bibinfo{pages}{629--659}.
%Type = Article
\bibitem[{Bico et~al.(2004)Bico, Roman, Moulin and Boudaoud}]{bico2004}
\bibinfo{author}{Bico, J.}, \bibinfo{author}{Roman, B.},
  \bibinfo{author}{Moulin, L.}, \bibinfo{author}{Boudaoud, A.},
  \bibinfo{year}{2004}.
\newblock \bibinfo{title}{Elastocapillary coalescence in wet hair}.
\newblock \bibinfo{journal}{Nature} \bibinfo{volume}{432},
  \bibinfo{pages}{690--690}.
%Type = Article
\bibitem[{Boudaoud and Cha\"ieb(2003)}]{bc2003}
\bibinfo{author}{Boudaoud, A.}, \bibinfo{author}{Cha\"ieb},
  \bibinfo{year}{2003}.
\newblock \bibinfo{title}{Mechanical phase diagram of shrinking cylindrical
  gels}.
\newblock \bibinfo{journal}{Phy. Rev. E} \bibinfo{volume}{68},
  \bibinfo{pages}{021801}.
%Type = Article
\bibitem[{Bush and Hu(2006)}]{bush2006}
\bibinfo{author}{Bush, J.W.}, \bibinfo{author}{Hu, D.L.}, \bibinfo{year}{2006}.
\newblock \bibinfo{title}{Walking on water: biolocomotion at the interface}.
\newblock \bibinfo{journal}{Annu. Rev. Fluid Mech.} \bibinfo{volume}{38},
  \bibinfo{pages}{339--369}.
%Type = Article
\bibitem[{Chater and Hutchinson(1984)}]{chater1984}
\bibinfo{author}{Chater, E.}, \bibinfo{author}{Hutchinson, J.},
  \bibinfo{year}{1984}.
\newblock \bibinfo{title}{On the propagation of bulges and buckles}.
\newblock \bibinfo{journal}{J. Appl. Mech} \bibinfo{volume}{51},
  \bibinfo{pages}{269--277}.
%Type = Article
\bibitem[{Chen et~al.(2012)Chen, Cai, Suo and Hayward}]{chen}
\bibinfo{author}{Chen, D.}, \bibinfo{author}{Cai, S.}, \bibinfo{author}{Suo,
  Z.}, \bibinfo{author}{Hayward, R.C.}, \bibinfo{year}{2012}.
\newblock \bibinfo{title}{Surface energy as a barrier to creasing of elastomer
  films: An elastic analogy to classical nucleation}.
\newblock \bibinfo{journal}{Phys. Rev. Lett} \bibinfo{volume}{109},
  \bibinfo{pages}{038001}.
%Type = Article
\bibitem[{Ciarletta(2011)}]{ciarletta2011}
\bibinfo{author}{Ciarletta, P.}, \bibinfo{year}{2011}.
\newblock \bibinfo{title}{Generating functions for volume-preserving
  transformations}.
\newblock \bibinfo{journal}{Int. J. Non-Linear Mech} \bibinfo{volume}{46},
  \bibinfo{pages}{1275--1279}.
%Type = Article
\bibitem[{Ciarletta(2014)}]{ciarletta2014wrinkle}
\bibinfo{author}{Ciarletta, P.}, \bibinfo{year}{2014}.
\newblock \bibinfo{title}{Wrinkle-to-fold transition in soft layers under
  equi-biaxial strain: A weakly nonlinear analysis}.
\newblock \bibinfo{journal}{J. Mech. Phys. Solids} \bibinfo{volume}{73},
  \bibinfo{pages}{118--133}.
%Type = Article
\bibitem[{Ciarletta and Ben~Amar(2012)}]{cb2012}
\bibinfo{author}{Ciarletta, P.}, \bibinfo{author}{Ben~Amar, M.},
  \bibinfo{year}{2012}.
\newblock \bibinfo{title}{Peristaltic patterns for swelling and shrinking of
  soft cylindrical gels}.
\newblock \bibinfo{journal}{Soft Matter} \bibinfo{volume}{6},
  \bibinfo{pages}{1760--1763}.
%Type = Article
\bibitem[{Datar et~al.(2019)Datar, Ameeramja, Bhat, Srivastava, Mishra, Bernal,
  Prost, Callan-Jones and Pullarkat}]{datar2019}
\bibinfo{author}{Datar, A.}, \bibinfo{author}{Ameeramja, J.},
  \bibinfo{author}{Bhat, A.}, \bibinfo{author}{Srivastava, R.},
  \bibinfo{author}{Mishra, A.}, \bibinfo{author}{Bernal, R.},
  \bibinfo{author}{Prost, J.}, \bibinfo{author}{Callan-Jones, A.},
  \bibinfo{author}{Pullarkat, P.A.}, \bibinfo{year}{2019}.
\newblock \bibinfo{title}{The roles of microtubules and membrane tension in
  axonal beading, retraction, and atrophy}.
\newblock \bibinfo{journal}{Biophys. J.} \bibinfo{volume}{117},
  \bibinfo{pages}{880--891}.
%Type = Book
\bibitem[{De~Gennes et~al.(2013)De~Gennes, Brochard-Wyart and
  Qu{\'e}r{\'e}}]{degennes2013}
\bibinfo{author}{De~Gennes, P.G.}, \bibinfo{author}{Brochard-Wyart, F.},
  \bibinfo{author}{Qu{\'e}r{\'e}, D.}, \bibinfo{year}{2013}.
\newblock \bibinfo{title}{Capillarity and wetting phenomena: drops, bubbles,
  pearls, waves}.
\newblock \bibinfo{publisher}{Springer Science \& Business Media}.
%Type = Article
\bibitem[{Dobyns et~al.(1993)Dobyns, Reiner, Carrozzo and
  Ledbetter}]{dobyns1993}
\bibinfo{author}{Dobyns, W.B.}, \bibinfo{author}{Reiner, O.},
  \bibinfo{author}{Carrozzo, R.}, \bibinfo{author}{Ledbetter, D.H.},
  \bibinfo{year}{1993}.
\newblock \bibinfo{title}{Lissencephaly: a human brain malformation associated
  with deletion of the lis1 gene located at chromosome 17p13}.
\newblock \bibinfo{journal}{Jama} \bibinfo{volume}{270},
  \bibinfo{pages}{2838--2842}.
%Type = Article
\bibitem[{Engstrom et~al.(2018)Engstrom, Zhang, Lawton, Joyner and
  Schwarz}]{engstrom2018}
\bibinfo{author}{Engstrom, T.}, \bibinfo{author}{Zhang, T.},
  \bibinfo{author}{Lawton, A.}, \bibinfo{author}{Joyner, A.},
  \bibinfo{author}{Schwarz, J.M.}, \bibinfo{year}{2018}.
\newblock \bibinfo{title}{Buckling without bending: a new paradigm in
  morphogenesis}.
\newblock \bibinfo{journal}{Phys. Rev. X} \bibinfo{volume}{8},
  \bibinfo{pages}{041053}.
%Type = Article
\bibitem[{Fu et~al.(2021)Fu, Jin and Goriely}]{FuST}
\bibinfo{author}{Fu, Y.}, \bibinfo{author}{Jin, L.}, \bibinfo{author}{Goriely,
  A.}, \bibinfo{year}{2021}.
\newblock \bibinfo{title}{Necking, beading, and bulging in soft elastic
  cylinders.}
\newblock \bibinfo{journal}{J. Mech. Phys. Solids} \bibinfo{volume}{147},
  \bibinfo{pages}{104250}.
%Type = Article
\bibitem[{Fu et~al.(2016)Fu, Liu and Francisco}]{fu2016localized}
\bibinfo{author}{Fu, Y.}, \bibinfo{author}{Liu, J.},
  \bibinfo{author}{Francisco, G.}, \bibinfo{year}{2016}.
\newblock \bibinfo{title}{Localized bulging in an inflated cylindrical tube of
  arbitrary thickness--the effect of bending stiffness}.
\newblock \bibinfo{journal}{J. Mech. Phys. Solids} \bibinfo{volume}{90},
  \bibinfo{pages}{45--60}.
%Type = Article
\bibitem[{Fu et~al.(2008)Fu, Pearce and Liu}]{fu2008}
\bibinfo{author}{Fu, Y.}, \bibinfo{author}{Pearce, S.}, \bibinfo{author}{Liu,
  K.K.}, \bibinfo{year}{2008}.
\newblock \bibinfo{title}{Post-bifurcation analysis of a thin-walled
  hyperelastic tube under inflation}.
\newblock \bibinfo{journal}{Int. J. Non-Lin. Mech} \bibinfo{volume}{43},
  \bibinfo{pages}{697--706}.
%Type = Article
\bibitem[{Giudici and Biggins(2020)}]{giudici2020}
\bibinfo{author}{Giudici, A.}, \bibinfo{author}{Biggins, J.S.},
  \bibinfo{year}{2020}.
\newblock \bibinfo{title}{Ballooning, bulging and necking: an exact solution
  for longitudinal phase separation in elastic systems near a critical point}.
\newblock \bibinfo{journal}{Phys. Rev. E} \bibinfo{volume}{102},
  \bibinfo{pages}{033007}.
%Type = Article
\bibitem[{Goriely et~al.(2015)Goriely, Geers, Holzapfel, Jayamohan,
  J{\'e}rusalem, Sivaloganathan, Squier, van Dommelen, Waters and
  Kuhl}]{goriely2015}
\bibinfo{author}{Goriely, A.}, \bibinfo{author}{Geers, M.G.},
  \bibinfo{author}{Holzapfel, G.A.}, \bibinfo{author}{Jayamohan, J.},
  \bibinfo{author}{J{\'e}rusalem, A.}, \bibinfo{author}{Sivaloganathan, S.},
  \bibinfo{author}{Squier, W.}, \bibinfo{author}{van Dommelen, J.A.},
  \bibinfo{author}{Waters, S.}, \bibinfo{author}{Kuhl, E.},
  \bibinfo{year}{2015}.
\newblock \bibinfo{title}{Mechanics of the brain: perspectives, challenges, and
  opportunities}.
\newblock \bibinfo{journal}{Biomech. Model. Mechanobiol} \bibinfo{volume}{14},
  \bibinfo{pages}{931--965}.
%Type = Article
\bibitem[{Haughton and Ogden(1979)}]{HO1979}
\bibinfo{author}{Haughton, D.}, \bibinfo{author}{Ogden, R.W.},
  \bibinfo{year}{1979}.
\newblock \bibinfo{title}{Bifurcation of inflated circular cylinders of elastic
  material under axial loading—ii. exact theory for thick-walled tubes}.
\newblock \bibinfo{journal}{J. Mech. Phys. Solids} \bibinfo{volume}{27},
  \bibinfo{pages}{489--512}.
%Type = Article
\bibitem[{Henann and Bertoldi(2014)}]{henann}
\bibinfo{author}{Henann, D.L.}, \bibinfo{author}{Bertoldi, K.},
  \bibinfo{year}{2014}.
\newblock \bibinfo{title}{Modeling of elasto-capillary phenomena}.
\newblock \bibinfo{journal}{Soft Matter} \bibinfo{volume}{10},
  \bibinfo{pages}{709--717}.
%Type = Book
\bibitem[{Iooss and Adelmeyer(1998)}]{Iooss}
\bibinfo{author}{Iooss, G.}, \bibinfo{author}{Adelmeyer, M.},
  \bibinfo{year}{1998}.
\newblock \bibinfo{title}{Topics in bifurcation theory and applications}.
  volume~\bibinfo{volume}{3}.
\newblock \bibinfo{publisher}{World Scientific}.
%Type = Article
\bibitem[{Kilinc et~al.(2009)Kilinc, Gallo and Barbee}]{kilinc}
\bibinfo{author}{Kilinc, D.}, \bibinfo{author}{Gallo, G.},
  \bibinfo{author}{Barbee, K.A.}, \bibinfo{year}{2009}.
\newblock \bibinfo{title}{Interactive image analysis programs for quantifying
  injury-induced axonal beading and microtubule disruption}.
\newblock \bibinfo{journal}{Comp. Meth. Progr. Biom} \bibinfo{volume}{95},
  \bibinfo{pages}{62--71}.
%Type = Article
\bibitem[{Kirchg{\"a}ssner(1982)}]{kirch}
\bibinfo{author}{Kirchg{\"a}ssner, K.}, \bibinfo{year}{1982}.
\newblock \bibinfo{title}{Wave-solutions of reversible systems and
  applications}.
\newblock \bibinfo{journal}{J. Diff. Eqns} \bibinfo{volume}{45},
  \bibinfo{pages}{113--127}.
%Type = Article
\bibitem[{Kyriakides and Yu-Chung(1991)}]{kyriakides1991}
\bibinfo{author}{Kyriakides, S.}, \bibinfo{author}{Yu-Chung, C.},
  \bibinfo{year}{1991}.
\newblock \bibinfo{title}{The initiation and propagation of a localized
  instability in an inflated elastic tube}.
\newblock \bibinfo{journal}{Int. J. Solids. Struct} \bibinfo{volume}{27},
  \bibinfo{pages}{1085--1111}.
%Type = Article
\bibitem[{Liu and Feng(2012)}]{lf2012}
\bibinfo{author}{Liu, J.L.}, \bibinfo{author}{Feng, X.Q.},
  \bibinfo{year}{2012}.
\newblock \bibinfo{title}{On elastocapillarity: A review}.
\newblock \bibinfo{journal}{Acta. Mech. Sin.} \bibinfo{volume}{28},
  \bibinfo{pages}{928--940}.
%Type = Article
\bibitem[{Mora et~al.(2011)Mora, Abkarian, Tabuteau and Pomeau}]{mora2011}
\bibinfo{author}{Mora, S.}, \bibinfo{author}{Abkarian, M.},
  \bibinfo{author}{Tabuteau, H.}, \bibinfo{author}{Pomeau, Y.},
  \bibinfo{year}{2011}.
\newblock \bibinfo{title}{Surface instability of soft solids under strain}.
\newblock \bibinfo{journal}{Soft matter} \bibinfo{volume}{7},
  \bibinfo{pages}{10612--10619}.
%Type = Article
\bibitem[{Mora et~al.(2010)Mora, Phou, Fromental, Pismen and Pomeau}]{mp2010}
\bibinfo{author}{Mora, S.}, \bibinfo{author}{Phou, T.},
  \bibinfo{author}{Fromental, J.M.}, \bibinfo{author}{Pismen, L.M.},
  \bibinfo{author}{Pomeau, Y.}, \bibinfo{year}{2010}.
\newblock \bibinfo{title}{Capillarity driven instability of a soft solid}.
\newblock \bibinfo{journal}{Phys. Rev. Lett} \bibinfo{volume}{105},
  \bibinfo{pages}{214301}.
%Type = Book
\bibitem[{Plateau(1873)}]{plateau1873}
\bibinfo{author}{Plateau, J.}, \bibinfo{year}{1873}.
\newblock \bibinfo{title}{Statique exp{\'e}rimentale et th{\'e}orique des
  liquides soumis aux seules forces mol{\'e}culaires}.
  volume~\bibinfo{volume}{2}.
\newblock \bibinfo{publisher}{Gauthier-Villars}.
%Type = Article
\bibitem[{Rayleigh(1892)}]{rayleigh1892}
\bibinfo{author}{Rayleigh, L.}, \bibinfo{year}{1892}.
\newblock \bibinfo{title}{On the instability of a cylinder of viscous liquid
  under capillary force}.
\newblock \bibinfo{journal}{Phil. Mag} \bibinfo{volume}{34},
  \bibinfo{pages}{145--154}.
%Type = Article
\bibitem[{Riccobelli and Bevilacqua(2020)}]{riccobelli2020}
\bibinfo{author}{Riccobelli, D.}, \bibinfo{author}{Bevilacqua, G.},
  \bibinfo{year}{2020}.
\newblock \bibinfo{title}{Surface tension controls the onset of gyrification in
  brain organoids}.
\newblock \bibinfo{journal}{J Mech Phys Solids} \bibinfo{volume}{134},
  \bibinfo{pages}{103745}.
%Type = Article
\bibitem[{Style et~al.(2017)Style, Jagota, Hui and Dufresne}]{style}
\bibinfo{author}{Style, R.W.}, \bibinfo{author}{Jagota, A.},
  \bibinfo{author}{Hui, C.Y.}, \bibinfo{author}{Dufresne, E.R.},
  \bibinfo{year}{2017}.
\newblock \bibinfo{title}{Elastocapillarity: Surface tension and the mechanics
  of soft solids}.
\newblock \bibinfo{journal}{Ann. Rev. Cond. Matter. Phys} \bibinfo{volume}{8},
  \bibinfo{pages}{99--118}.
%Type = Article
\bibitem[{Taffetani and Ciarletta(2015)}]{taffetani}
\bibinfo{author}{Taffetani, M.}, \bibinfo{author}{Ciarletta, P.},
  \bibinfo{year}{2015}.
\newblock \bibinfo{title}{Beading instability in soft cylindrical gels with
  capillary energy: weakly non-linear analysis and numerical simulations}.
\newblock \bibinfo{journal}{J. Mech. Phys. Solids} \bibinfo{volume}{81},
  \bibinfo{pages}{91--120}.
%Type = Article
\bibitem[{Tanaka et~al.(1992)Tanaka, Tomita, Takasu, Hayashi and
  Nishi}]{tanaka1992}
\bibinfo{author}{Tanaka, H.}, \bibinfo{author}{Tomita, H.},
  \bibinfo{author}{Takasu, A.}, \bibinfo{author}{Hayashi, T.},
  \bibinfo{author}{Nishi, T.}, \bibinfo{year}{1992}.
\newblock \bibinfo{title}{Morphological and kinetic evolution of surface
  patterns in gels during the swelling process: Evidence of dynamic pattern
  ordering}.
\newblock \bibinfo{journal}{Phys. Rev. Lett} \bibinfo{volume}{68},
  \bibinfo{pages}{2794}.
%Type = Article
\bibitem[{Wang(2020)}]{liuwang}
\bibinfo{author}{Wang, L.}, \bibinfo{year}{2020}.
\newblock \bibinfo{title}{Axisymmetric instability of soft elastic tubes under
  axial load and surface tension}.
\newblock \bibinfo{journal}{Int. J. Solids. Struct.} \bibinfo{volume}{191},
  \bibinfo{pages}{341--350}.
%Type = Article
\bibitem[{Wilkes(1955)}]{wi1955}
\bibinfo{author}{Wilkes, E.}, \bibinfo{year}{1955}.
\newblock \bibinfo{title}{On the stability of a circular tube under end
  thrust}.
\newblock \bibinfo{journal}{Q. J. Mech. Appl. Math} \bibinfo{volume}{8},
  \bibinfo{pages}{88--100}.
%Type = Article
\bibitem[{Wineman(2005)}]{wineman2005}
\bibinfo{author}{Wineman, A.}, \bibinfo{year}{2005}.
\newblock \bibinfo{title}{Some results for generalized neo-hookean elastic
  materials}.
\newblock \bibinfo{journal}{Int. J. Non-Lin. Mech.} \bibinfo{volume}{40},
  \bibinfo{pages}{271--279}.
%Type = Manual
\bibitem[{{Wolfram Research Inc.}(2019)}]{wo2019}
\bibinfo{author}{{Wolfram Research Inc.}}, \bibinfo{year}{2019}.
\newblock \bibinfo{title}{Mathematica 12.0}.
\newblock \bibinfo{note}{Wolfram Research Inc, Champaign, IL}.
%Type = Article
\bibitem[{Xuan and Biggins(2016)}]{xuan2016}
\bibinfo{author}{Xuan, C.}, \bibinfo{author}{Biggins, J.},
  \bibinfo{year}{2016}.
\newblock \bibinfo{title}{Finite-wavelength surface-tension-driven
  instabilities in soft solids, including instability in a cylindrical channel
  through an elastic solid}.
\newblock \bibinfo{journal}{Phys. Rev. E} \bibinfo{volume}{94},
  \bibinfo{pages}{023107}.
%Type = Article
\bibitem[{Xuan and Biggins(2017)}]{xuan2017}
\bibinfo{author}{Xuan, C.}, \bibinfo{author}{Biggins, J.},
  \bibinfo{year}{2017}.
\newblock \bibinfo{title}{Plateau-rayleigh instability in solids is a simple
  phase separation}.
\newblock \bibinfo{journal}{Phys. Rev. E} \bibinfo{volume}{95},
  \bibinfo{pages}{053106}.
%Type = Article
\bibitem[{Zhou et~al.(2018)Zhou, Wang, Li and Fu}]{zhou2018}
\bibinfo{author}{Zhou, L.}, \bibinfo{author}{Wang, S.}, \bibinfo{author}{Li,
  L.}, \bibinfo{author}{Fu, Y.}, \bibinfo{year}{2018}.
\newblock \bibinfo{title}{An evaluation of the gent and gent-gent material
  models using inflation of a plane membrane}.
\newblock \bibinfo{journal}{Int. J. Mech. Sci.} \bibinfo{volume}{146},
  \bibinfo{pages}{39--48}.

\end{thebibliography}

\end{document}